\documentclass[11pt,letterpaper]{article}
\pdfoutput=1
\usepackage[nocompress]{cite}
\usepackage{graphicx}
\usepackage{url}

\title{Good Colour Maps: How to Design Them} 

\author{Peter Kovesi\\
Centre for Exploration Targeting\\
School of Earth and Environment \\
The University of Western Australia\\
Crawley, Western Australia, 6009\\
\texttt{peter.kovesi@uwa.edu.au}}
\date{September 2015}

\begin{document} 
\maketitle

% The paper headers
%\markboth{Peter Kovesi}{Good Colour Maps: How to Design Them}
\begin{abstract}
Many colour maps provided by vendors have highly uneven perceptual
contrast over their range.  It is not uncommon for colour maps to have
perceptual flat spots that can hide a feature as large as one tenth of
the total data range.  Colour maps may also have perceptual
discontinuities that induce the appearance of false features.
Previous work in the design of perceptually uniform colour maps has
mostly failed to recognise that CIELAB space is only designed to be
perceptually uniform at very low spatial frequencies.  The most
important factor in designing a colour map is to ensure that the
magnitude of the incremental change in perceptual lightness of the
colours is uniform.  The specific requirements for linear, diverging,
rainbow and cyclic colour maps are developed in detail.  To support
this work two test images for evaluating colour maps are presented.
The use of colour maps in combination with relief shading is
considered and the conditions under which colour can enhance or
disrupt relief shading are identified.  Finally, a set of new basis
colours for the construction of ternary images are presented.  Unlike
the RGB primaries these basis colours produce images whereby the
salience of structures are consistent irrespective of the assignment
of basis colours to data channels.
\end{abstract}

\section{Introduction}
A colour map can be thought of as a line or curve drawn through a
three dimensional colour space. Individual data values are mapped to
positions along this line which, in turn, allows them to be mapped to
a colour.  For a colour map to be effective it should allow the
structure and form of the data to be seen and, ideally, allow the
communication of metric information in the data (its values) as
well~\cite{Ware1988}.  Achieving a good representation of the
structure of the data is primarily achieved by ensuring that the
perceptual contrast that occurs as one moves along the colour map path
in colour space is close to uniform and that the colours in the map
follow an intuitive perceptual ordering.  Communicating accurate
metric information via colours in the map is difficult and is
inevitably compromised due to simultaneous contrast and chromatic
contrast effects.  Typical viewing conditions are not controlled and
most displays are not calibrated, the best one can reasonably hope to
communicate is some qualitative metric information. However, there is
no good reason why structure and form information should be
compromised in any way.

Unfortunately many widely used colour maps
\footnote{GIS environments may make the distinction between
  \emph{colour maps} and \emph{colour ramps} whereby a colour map is
  obtained by employing an algorithm such as histogram equalisation
  or linear stretching to map data values to colours on a colour
  ramp.  The form of the colour `ramp' is arbitrary and is not
  necessarily a ramp.  In GIS terms this paper is about the design of
  colour ramps.}
provided by vendors have
highly uneven perceptual contrast.  Colour maps may have points of
locally high colour contrast leading to the perception of false
anomalies in your data when there is none.  Conversely colour maps may
also have `flat spots' of low perceptual contrast that prevent you
from seeing features in the data.

\begin{figure}
\centering
\includegraphics[width=11.cm]{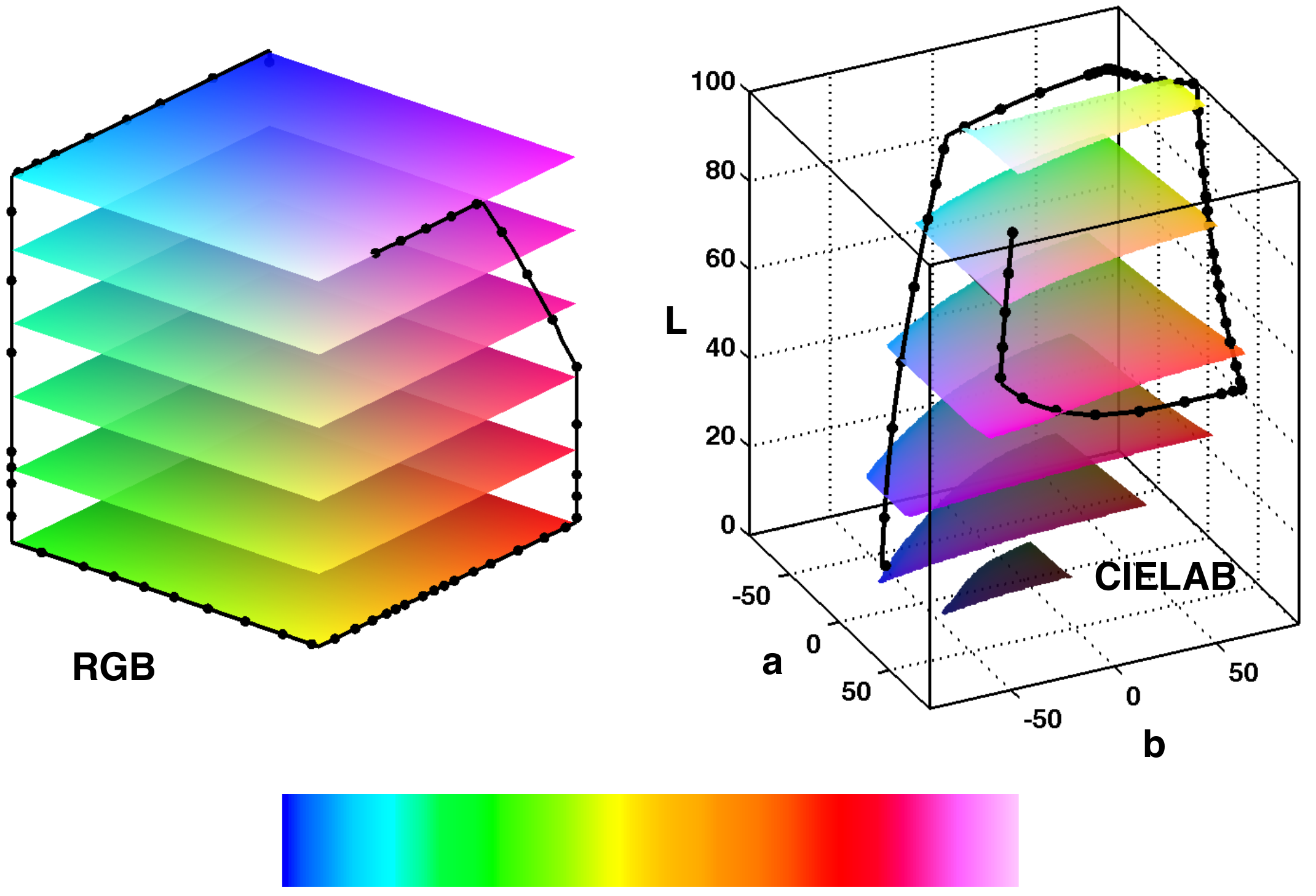}
\caption{A colour map and its path through RGB and CIELAB colour
  spaces.}
  \label{fig:geosoft_rgb_lab}
\end{figure}

In many cases these problems arise because the colour maps have been
designed as piecewise linear paths through RGB space. However, this
colour space is not the best in which to design or analyse a colour
map.  It is better to use a colour space such as CIELAB which has been
designed to be perceptually uniform whereby distances between points
in the 3D colour space are intended to closely correspond to human
perception of colour difference.  CIELAB space represents colour in
term of lightness, varying from 0 to 100, and $a$ and $b$ components
nominally representing the red-green and yellow-blue opponent channels
respectively. The vertical axis through this space at $a = 0,\ b = 0$
corresponds to the grey scale. The distance away from this axis
corresponds to chroma.  Note that while CIELAB is designed to be
perceptually linear this is only true for visual stimuli at very low
spatial frequencies, this is not widely recognised but has a
significant bearing on colour map design.  This will be discussed
further in Section~\ref{sec:lightnessgradient}.

Figure~\ref{fig:geosoft_rgb_lab} shows a typical rainbow style colour
map along with its path through the RGB and CIELAB colour spaces. Its
straight line construction in RGB space is evident.  The spacing of
the dots along the paths in the respective colour spaces is
proportional to the spacing of adjacent values in the colour map.
Notice the uneven spacing within the CIELAB colour space. The
clustering of points in the green and red regions produce perceptual
flat spots in the colour map.  The extended section of near constant
lightness between cyan and yellow exacerbates the green flat spot.
The kinks and uneven point spacing along the curve at cyan, yellow and
red produce the false anomalies seen at these points.  The reversal of
lightness gradients at yellow and red also contribute to this.  

It should be emphasised that this paper is concerned with the design
of colour maps for rendering data that varies over a continuous range
such as geophysical exploration images or medical imagery.  For colour
maps suited to the display of data consisting of a limited number of
categorical values it is suggested that you refer to the work by
Brewer~\cite{Brewer1994a,Brewer1994b,Brewer1997,Brewer2013}. However,
while the emphasis of Brewer's work is mainly directed towards
cartographic applications many of her design techniques and principles
are also relevant here.

\subsection{Evaluating Colour Maps}
\label{sub:evaluatingmaps}

\begin{figure}
\centering
\includegraphics[width=9cm]{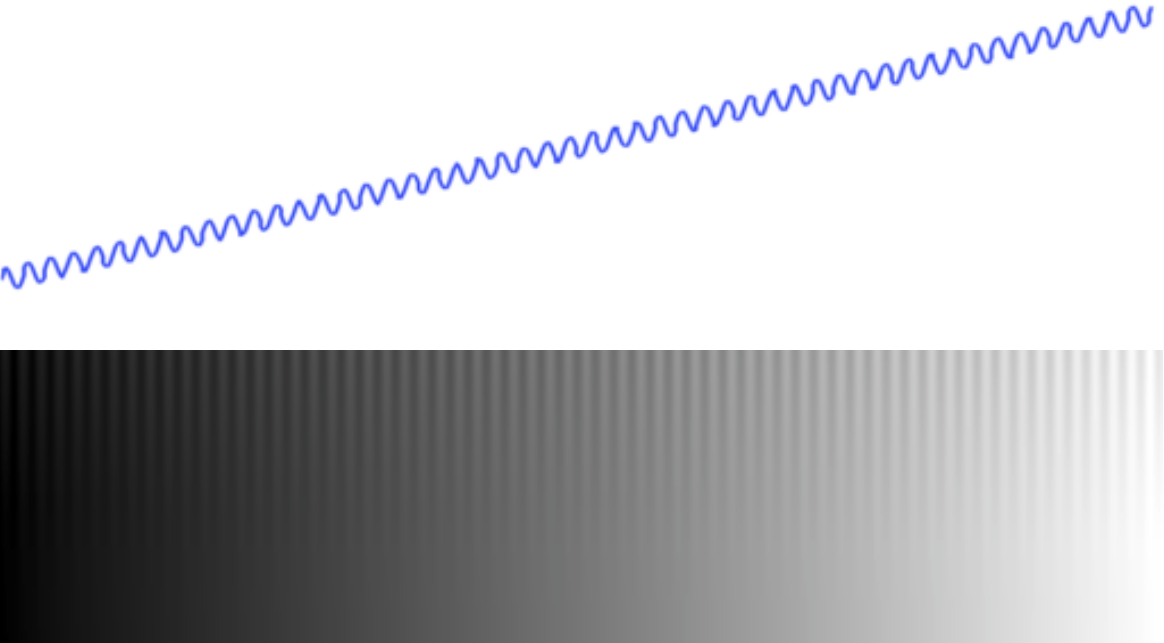}
\caption{The colour map test image and its cross section.}
  \label{fig:lineartestimage}
\end{figure}

Possibly a contributing factor to the proliferation of poor colour
maps has been the absence of a simple test image that allows colour
maps to be evaluated.  The test image shown in
Figure~\ref{fig:lineartestimage} attempts to remedy this.  Its design
is inspired by the sinusoidal gratings used for psychophysical
contrast sensitivity
tests~\cite{OlzakThomas1986,VanNesBouman1967,WatsonEtAl1983}.  It
consists of a sine wave superimposed on a ramp function, this provides
a set of constant magnitude features presented at different offsets.
The spatial frequency of the sine wave is chosen to lie in the range
at which the human eye is most sensitive\footnote{Note the figures in this paper present the test image slightly
 smaller than its designed size.}, 
and its amplitude is set so that the range from peak to
trough represents a series of features that are 10\% of the total data
range.  The amplitude of the sine wave is modulated from its full
value at the top of the image to zero at the bottom.  If the colour
map has uniform perceptual contrast the sine wave should be uniformly
visible across the full width of the image.  At the very bottom of the
image, where the sine wave amplitude is zero, we just have a linear
ramp which simply reproduces the colour map.  We should not perceive
any identifiable features across the ramp.

Of course viewing a colour map on this test image cannot replace a
detailed psychophysical evaluation but it allows immediate recognition
of any serious faults in the map (or your display monitor).
Recognising that most images are viewed in uncontrolled conditions on
uncalibrated displays this is all that is necessary except for the most
exacting of applications.  This test image immediately reveals the
perceptual flat spots and false features in the vendor colour maps
shown in Figure~\ref{fig:vendorproblem}.  For details on the design of
the test image see Appendix~\ref{appendix:test_image}.

\begin{figure}
\centering
\includegraphics[width=9cm]{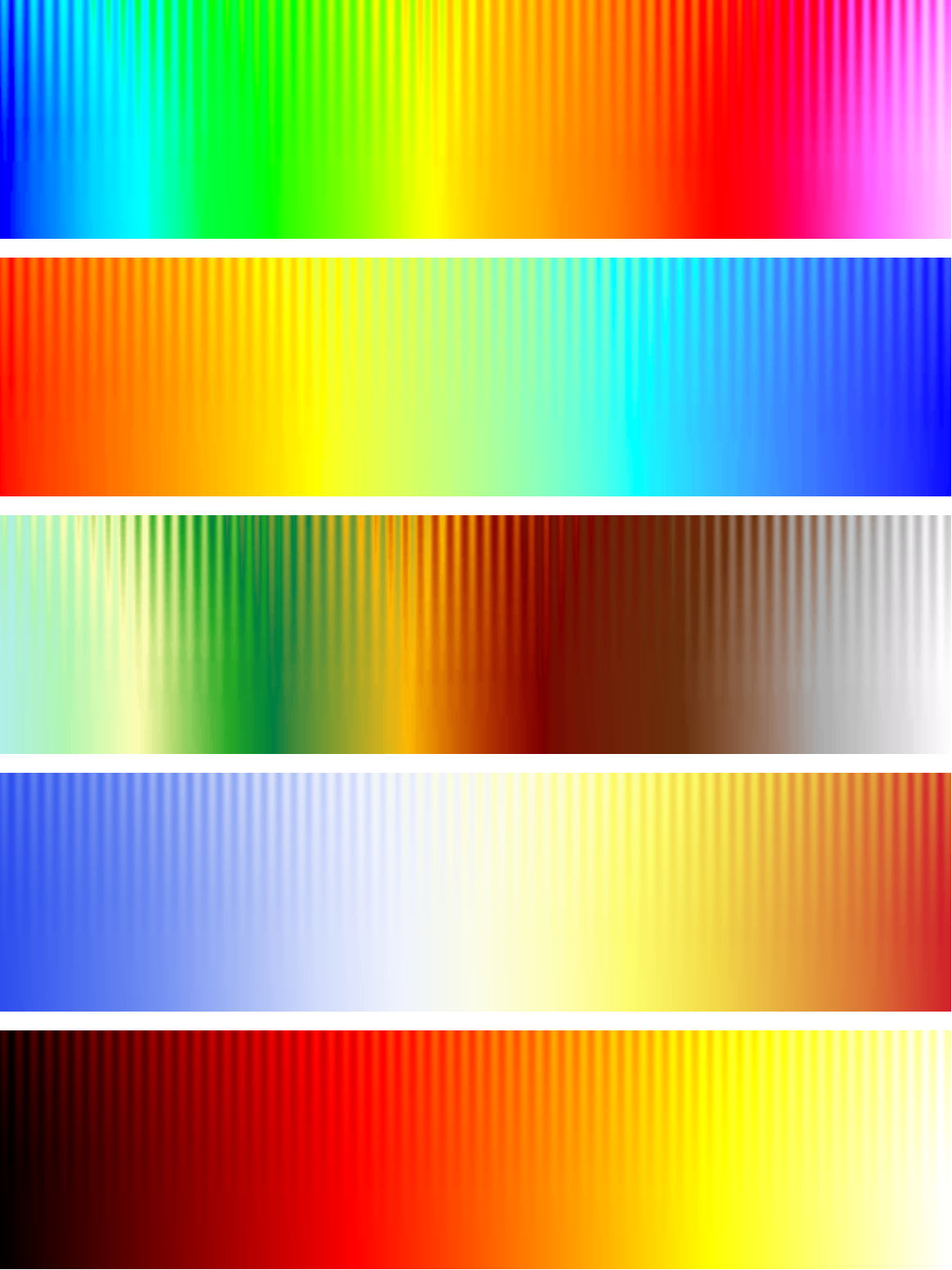}
\caption{ The test image rendered with a selection of vendor colour
  maps. The vendors represented here include MathWorks, Geosoft, ESRI
  and Wolfram.}
  \label{fig:vendorproblem}
\end{figure}

\section{The Importance of Lightness}
\label{sec:lightnessgradient}
In order to achieve uniform perceptual contrast in a colour map an
initial approach might be to set the spacing of colour map points
along a path through colour space according to a colour difference
formula such as CIE76~\cite{CIE76}, which corresponds to the Euclidean
distance in CIELAB space, or CIEDE2000~\cite{CIE2001,LuoCuiRigg2001}.
This approach has been used by numerous workers including
Pizer~\cite{Pizer1981}, Tajima~\cite{Tajima1983}, Robertson and
O'Callaghan~\cite{RobertsonOCallaghan1986}, Levkowitz and
Herman~\cite{LevkowitzHerman1992}, Rogowitz
et~al.~\cite{RogowitzKalvinPelahCohen1999} and
Moreland~\cite{Moreland2009}.  In all these cases CIE76, or the
Euclidean distance in CIELUV space, was used as the perceptual
contrast measure.

However, the problem with this approach is that these colour difference
formulas were derived from experiments where human subjects were asked
to compare large isolated patches of colour that subtend a significant
field of view.  Initially these difference formulas were developed
from the CIE~1931 $2^{\circ}$ Standard
Observer~\cite{Wright1928,Guild1931,CIE1931} and then subsequently the
CIE~1964 $10^{\circ}$ Standard
Observer~\cite{StilesBurch1958,Speranskaya1959,Fairchild2013}.  Noting
that a 10mm screen object viewed at a distance of 600mm subtends about
1 degree of visual angle we can see that these colour difference
formulas are only valid for spatial scales that are very much larger
than what we are typically seeking to resolve in an image.

At fine spatial scales acuity performance on chromatic gratings is
significantly lower than it is for achromatic luminance
gratings~\cite{PoirsonWandell1993,Mullen1985,LennieDZmura1988}.
Mullen~\cite{Mullen1985} indicates that the contrast sensitivity
function of red-green and blue-yellow gratings is characteristic of a
low-pass filter.  Acuity performance starts to decrease significantly
for spatial frequencies greater than about 3 cycles/degree with
resolution ultimately failing at about 11-12 cycles/degree.  In
addition to spatial frequency effects at small fields of view, below
0.5 degrees of viewing angle, there is a severe loss of colour
discrimination and ultimately below about 0.3 degrees an observer with
normal trichromatic vision becomes
dichromatic~\cite{ThomsonWright1947}.  Thus, when one talks about
CIELAB space being perceptually uniform this is really only the case
for very low spatial frequencies.  In acknowledgment of this a spatial
extension to CIELAB, S-CIELAB, was developed by Zhang and
Wandell~\cite{ZhangWandell1996} by introducing a pre-processing
spatial filering step before computing the CIE76 colour difference.
The spatial filters being designed to approximate the human contrast
sensitivity function for achromatic and chromatic signals.  Johnson
and Fairchild~\cite{JohnsonFairchild2003} subsequently applied these
ideas to form a spatial extension to CIEDE2000.  At high spatial
frequencies these modified measures become dominated by lightness
differences.  The effect of spatial scale on the perception of colour
difference has been studied by Stone~\cite{Stone2012} for the design
of graphs and charts.  She notes that while large patches of different
isoluminant colours can be readily distinguished this is not the case
when the same colours are used at fine spatial scales, say for line
graphs or scatter plots.  For these kinds of charts legibility at fine
spatial scales becomes dominated by luminance contrast.  Subsequent
work by Stone et al.~\cite{StoneEtAl2014} describes a preliminary
attempt to determine a scale dependent, non-uniform rescaling of
CIELAB space for colour difference calculations using data from crowd
sourced experiments.

Thus, for the purposes of colour map design, where we are interested
in the ability to resolve fine structures within images, the
perceptual contrast between colours is dominated by the difference in
the lightness of the colours.  Any difference in hue or
chroma/saturation is relatively unimportant.

\begin{figure}
\centering
\includegraphics[width=12.cm]{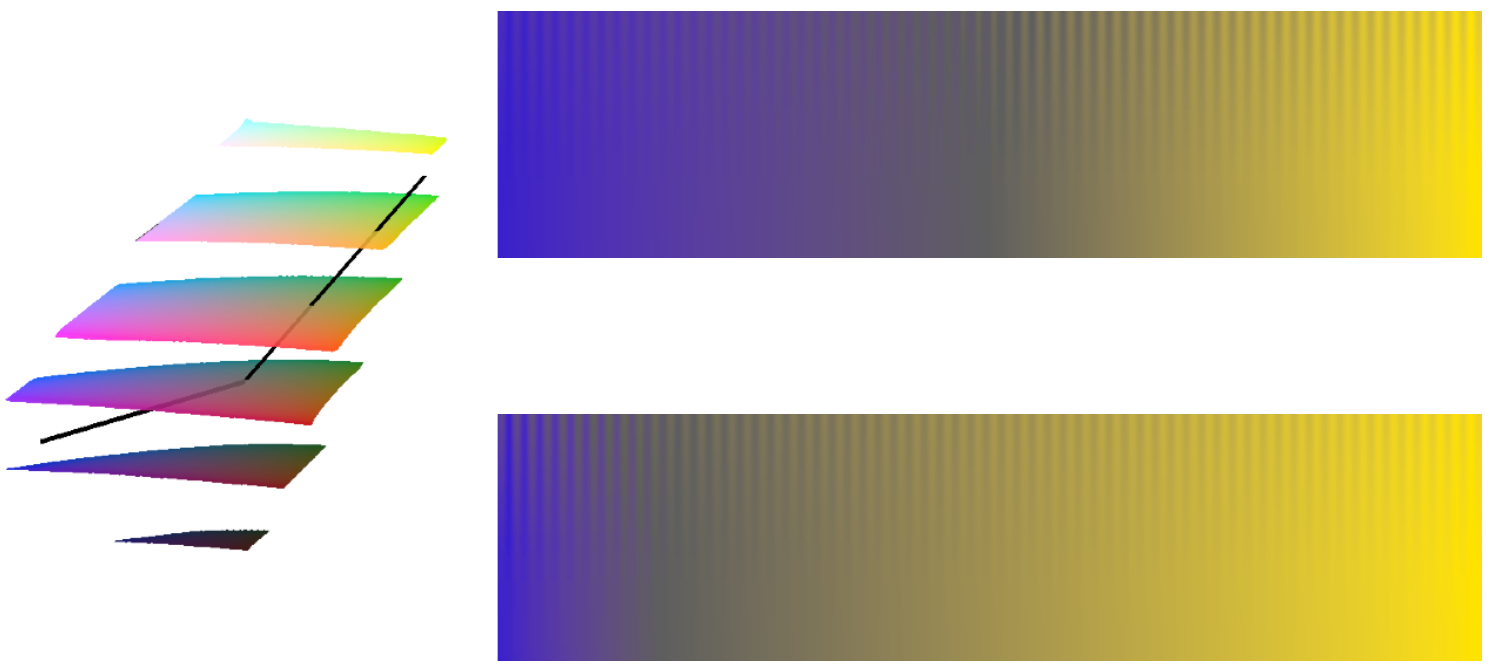}
\caption{Two colour maps formed from the same path through colour
  space shown on the left.  The top one is constructed from points 
  equispaced in CIELAB space, the lower one is constructed from points 
  equispaced just in terms of CIELAB lightness.}
\label{fig:lightness_importance}
\end{figure}

Figure~\ref{fig:lightness_importance} provides an illustration of the
importance of lightness gradient with respect to position along the
colour map. Two colour maps are constructed from the same path through
CIELAB space.  The path consists of two line segments of equal length
but of different slope through the colour space.  One colour map is
generated by selecting points equispaced along the path.  This
corresponds to points having equal colour contrast under the CIE76
equation.  The other map is generated by selecting points at equal
increments of CIELAB lightness.
% lightnessimportance.m

When rendered on the test image the colour map formed from equispaced
points can be divided into two sections corresponding to the two
segments of different slope.  The right half of the colour map
provides good feature contrast.  The left half renders the test image
poorly with the feature contrast very much reduced as a result of the
reduced lightness gradient across this part of the map.  On the other
hand the colour map formed from points of equal increment in lightness
value renders the test image well.  The sine wave pattern is seen
uniformly across the full width of the test image.  Note the blue
section of the colour map is compressed because there is a small
change in lightness across this segment.  Thus, the number of colour
map points on this section of the path are correspondingly reduced.

\begin{figure}
\centering
\includegraphics[width=9cm]{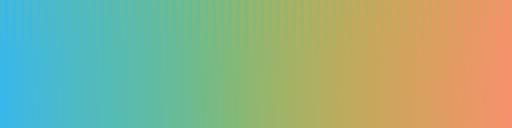}
\caption{The test image rendered with an isoluminant colour map.}
\label{fig:isoluminant59}
\end{figure}

The importance of the lightness gradient in a colour map is made very
evident when one constructs a colour map of constant lightness.
Figure~\ref{fig:isoluminant59} shows a map generated from equispaced
points on a curve through CIELAB space at a lightness level of 70.
Notice how the sine wave pattern in the test image is almost
impossible to discern.  At first sight such a colour map would seem to
be a poor choice for displaying data.  However, as will be shown
later, constant lightness and low contrast colour maps can be useful
when displaying data with relief shading.

\section{Prior Work}
\label{sec:priorwork}

As mentioned above there have been many attempts to generate
perceptually uniform colour maps using the Euclidean distance in
CIELAB or CIELUV space as a measure of perceptual
contrast~\cite{Pizer1981,Tajima1983,RobertsonOCallaghan1986,
LevkowitzHerman1992,RogowitzKalvinPelahCohen1999,Moreland2009}.
In general this work failed to recognize that the perceptual
uniformity of CIELAB is not valid at fine spatial scales and this has
resulted in inconsistent success in generating good colour maps.  For
example Levkowitz and Herman~\cite{LevkowitzHerman1992} constructed a
colour map path that was designed to maximise the traversed CIELUV
distance while also maintaining a colour ordering.  However, they
reported that this map was less effective than a linearised grey scale
when evaluated on medical images.  This was despite their optimised
map traversing a CIELUV distance six times that of the grey scale map.
Rogowitz et~al.~\cite{RogowitzKalvinPelahCohen1999}, in their search
for good colour maps for representing magnitude information,
constructed color maps which traced carefully controlled paths through
Munsell and CIELAB color spaces.  They concluded that luminance and
saturation were good candidates for representing magnitude and that
hue based maps performed poorly.

Rogowitz and Treinish~\cite{RogowitzTreinish1988} recognised that the
chromatic and achromatic responses of the eye have very different
characteristics with respect to spatial frequency.  They make the
suggestion that low frequency information in data can be mapped to
colour saturation while dedicating luminance for encoding high
frequency information.  They point out a number of problems with
rainbow maps, with colour ordering being confused and viewers tending
to partition images into uneven bands.  Ware~\cite{Ware1988} makes the
distinction between the need to identify the data's form and its
metric information.  Simultaneous contrast and chromatic contrast
effects makes accurate metric information from lightness or colour
difficult but he concludes that if you wish to read metric quantities
with a colour key then a rainbow like map works well.  For maximal
form information he suggests a grey scale should be used.  Where both
are required a sequence that increases monotonically in luminance
while cycling through a range of hues is suggested.

Kindlmann et~al.\cite{KindlmannReinhardCreem2002} devised a novel way
of constructing perceptual colour maps by exploiting our ability to
recognise whether a face is being presented as a negative or positive
image.  This was used to perform luminance matching between colours
and a grey scale.  From this they were able to demonstrate the
generation of an isoluminant colour map and a map of monotonically
increasing lightness.  Rogowitz and Kalvin~\cite{RogowitzKalvin2001}
used face images for the evaluation of colour maps.  They mapped
various color scales onto the intensity values of a face and asked
viewers to rate the images for their `naturalness'.  They concluded
that monotonically increasing luminance in a colour map was important,
and that rainbow maps performed badly.  This conclusion is perhaps not
surprising given that any map with non-monotonically increasing
luminance will disrupt the shading pattern on a face, making it look
unnatural.

Kalvin~\cite{Kalvin2002} makes the point that, as well as perceptual
uniformity, it is important that a colour map is perceptually ordered.
He constructs a directed graph on the 12 edges of the RGB cube and
suggests that if a colour map path traverses any section of the graph
in the specified direction then it will have an appropriate perceptual
ordering.  McNames~\cite{McNames2006} proposed a colour map designed
to reproduce well in both colour and grey scale. It is based on a
spiral path through RGB space. The map has a monotonically increasing
lightness scale but the perceptual smoothness of its colour variations
is not ideal.

In relation to diverging colour maps Spence and
Efendov~\cite{SpenceEfendov2001} assessed the ability of human
subjects to discriminate isoluminant targets in an image display.
They set up isoluminant bipolar/diverging colour maps that passed from
one colour through grey to another colour at a lightness level of 85.
Colours that differed in hue angle by 180 degrees in CIELAB space were
not necessarily the best performers with hue angle differences of 90
degrees often performing well.  Pairs of colours with non-negative
CIELAB $b$ components (on the yellow side of CIELAB rather than blue)
tended to do well. However, these result were for isoluminant targets
and it likely that results for maps varying in luminance would be
different.

More recently the importance of the lightness gradient in a colour map
is noted by Niccoli~\cite{NiccoliBlog2012}.  In designing his
perceptual colour maps he ensures the lightness profiles are linear,
or follow a cube law to match Steven's power law~\cite{Stevens1957}.
However his reasoning for using a cube law is not clear given that
CIELAB lightness is intended to be perceptually linear.

\section{Colour Map Design}
\label{sec:colour_map_design}

The initial step is to design a path through colour space that one wishes the 
colour map to traverse.  
The colour maps presented here are constructed by specifying control
points in CIELAB space and then fitting a 1st or 2nd order B-spline
through them to form the colour map path.  A 1st order spline forms a
linear path between the control points and a 2nd order one uses
quadratic basis functions to form the path.  B-splines provide a
general and flexible means of defining colour map paths. This is
important because colour map paths often have to be carefully hand
crafted in order to achieve a desired perceptual outcome within the
sometimes awkward constraints of the colour space gamut.  Having
formed a colour map path one then needs to determine a set of locations along
the path that are equispaced in terms of perceptual contrast to form
the final colour map.

% Talk about charting locations in CIELAB space that you want the colour
% map to traverse and use these as control points for the B-spline

The technique used here for obtaining N colour map values of equal
perceptual contrast along a colour map path is similar that originally
used by Pizer~\cite{Pizer1981} for linearizing the perceptual contrast 
on CRT monitors.  The process is analogous to performing histogram 
equalisation of an image whereby the  cumulative distribution of image 
values is used as a remapping function to obtain a uniform distribution 
of grey values.  In our application, for colour maps, we use the cumulative 
sum of perceptual  contrast  differences along the map as the remapping 
function for equalising the perceptual contrast.

\begin{figure}
\centering
\includegraphics[width=10cm]{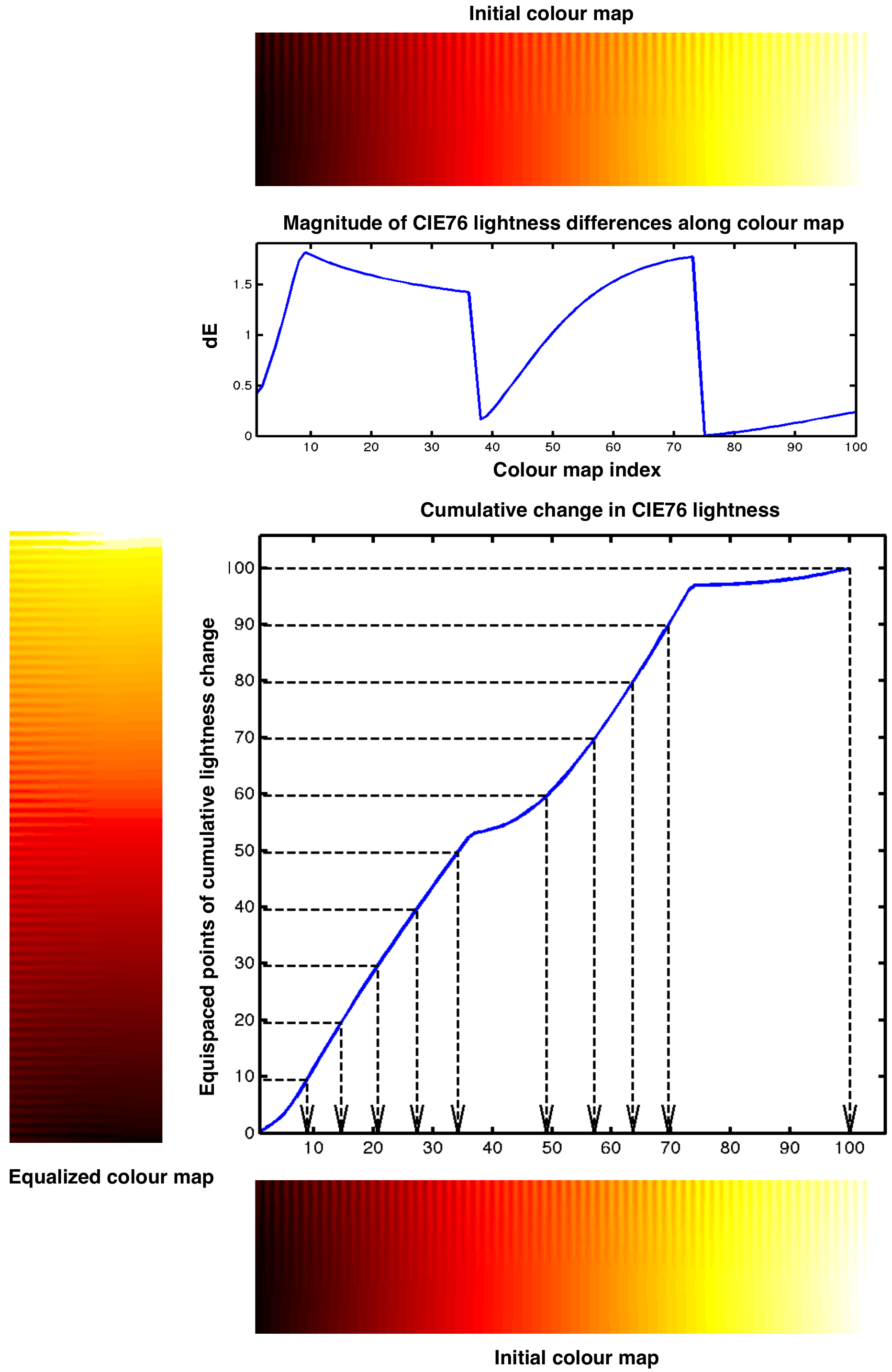}
\caption{The perceptual contrast equalization process: At the top is
  the initial colour map and a plot of the lightness differences
  between successive colours along the map.  Note how low dE values
  correspond to perceptual flat spots in the map.  Below is the
  cumulative change in lightness. This is used as the mapping function
  that takes equispaced points in cumulative lightness change, which
  are used to form the equalized colour map on the left, and maps them
  to their source locations in the initial map, which is shown again
  at the bottom.  }
\label{fig:equalizationprocess}
\end{figure}

First, an initial colour map is generated by evaluating N points at
equal spline parameter increments along the path.  The perceptual
differences between successive colour map entries are then computed.
For most colour maps this will be simply the magnitude of lightness
differences between successive colour map entries.  However, for
isoluminant, or low lightness contrast colour maps, the CIE76 formula
would be used.  From this a cumulative sum of the contrast differences
along the colour map is formed.  The total cumulative contrast change
is then divided into N equispaced values and a reverse mapping back to
the spline parameters required to obtain these equispaced contrast
values is obtained via linear interpolation of the cumulative contrast
curve.  These new remapped locations form the final colour map.  In
practice this process is repeated recursively on its own output. This
helps overcome any approximations induced by using linear
interpolation to estimate the locations of equal perceptual contrast
in the reverse mapping. This is mainly an issue for colour maps with
only a few entries.  Of course this contrast equalisation process can
be applied to existing colour maps as well as being used for the
design of new maps. The overall process is illustrated in
Figure~\ref{fig:equalizationprocess} using MATLAB's `hot' colour
map~\cite{MATLABhot2014} as the input map requiring perceptual
contrast equalization.

% generateequaliseplot.m

For most colour maps where the perceptual contrast is dominated by
lightness variations it is important to ensure that the colour map
path does not have extended segments with little or no lightness
gradient. If this is the case then when the path is sampled for points
of equidistant lightness change one may end up with undesirable hue
and/or chroma discontinuities in the colour map. Any attempt to remove
the hue or chroma discontinuities by incorporating the CIE76 colour
difference formula in the perceptual contrast equalization process is
complicated by the fact that one should be mindful of the different
characteristics of the eye's sensitivity to contrast in luminance and
chromaticity. Luminance contrast sensitivity is band-pass in nature
whereas chromatic contrast sensitivity is low-pass in nature~\cite{Mullen1985}. Thus,
the perceptual contrast across a colour map is inescapably a function
of spatial scale. However, to minimise the effect of scale a colour
map path should be dominated either by lightness changes or by
chromatic changes, but not a mixture of both. We want to avoid any
perceptual contrast equalization process requiring luminance contrast
to be played off against chromatic contrast over different sections of
the colour map. Under this situation one is likely to produce a
map that is especially scale dependent with regard to perceptual
contrast.

\subsection{CIE76 or CIEDE2000?}

Various limitations of the CIE76 colour difference measure ultimately
led to the development of the CIEDE2000 formula which incorporates
correction factors that are applied to the differences in lightness,
chroma and hue~\cite{SharmaWuDalal2005}. In seeking to equalize the
perceptual contrast across a colour map the question arises as to
which formula one should use.
%A basic evaluation of the two formulas
%is provided in Appendix~\ref{sec:cie76_ciede2000}. 
A basic evaluation of the two formulas reveals that the lightness
correction that is incorporated in CIEDE2000 is possibly useful but
its effect is barely noticeable.  However, the CIEDE2000 chroma
correction is considerably more significant~\cite{SharmaWuDalal2005}.
This correction noticeably emphasizes the contrast in high chroma
regions at the expense of the low chroma sections.  At the high
spatial frequencies of interest to us this correction proves to be
quite inappropriate. Thus, in the absence of any alternative it would
appear that CIE76 is the most appropriate colour difference formula to
use. Though, for most cases, we are only interested in its lightness
component. Ultimately one should remember that neither colour
difference formula was designed for the spatial frequencies that we
are interested in.

\subsection{A Taxonomy of Colour Maps}
\label{sub:taxonomy}

Colour maps can be organized according to the following attributes:
linear, diverging, rainbow, cyclic, and isoluminant.
\\ \\ \includegraphics[width=3.5cm,  height=0.5cm]{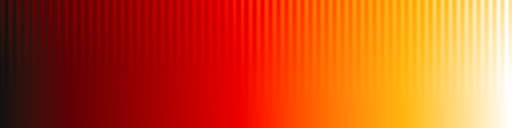}\\ 
Linear colour maps have colour lightness values that increase or
decrease linearly over the colour map's range and are intended for
general use.  These colour maps can be considered a subset of what are
know as {\em sequential} maps~\cite{Brewer1994a,Brewer1994b}.  The
term linear is used here to emphasise the fact that we are concerned
with maps for which the lightness gradient is constant.
\\ \\ \includegraphics[width=3.5cm,  height=0.5cm]{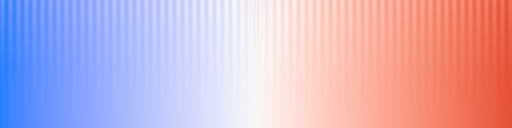}\\ 
Diverging colour maps follow some pattern of symmetry about their
centre.  They are suitable where the data has a specific reference
value and we are interested in differentiating values that lie above,
or below, this value.
\\ \\ \includegraphics[width=3.5cm,  height=0.5cm]{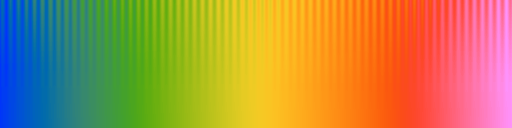}\\ 
Rainbow colour maps, which nominally follow some representation of the
spectrum, have well documented
shortcomings~\cite{Brewer1994a,Brewer1997,RogowitzTreinish1988,BorlandTaylor2007}.  However,
rainbow colour maps are ubiquitous and are unlikely to go away.
Accordingly they warrant a category of their own.
\\ \\ \includegraphics[width=1.5cm,  height=1.5cm]{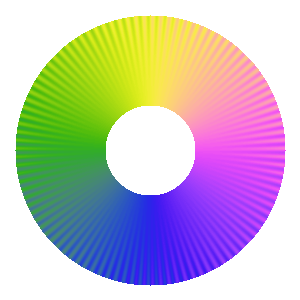}\\ 
Cyclic colour maps have colours that are matched at each end with
first order continuity.  They are intended for the presentation of
orientation values or angular phase data.
\\ \\ \includegraphics[width=3.5cm,  height=0.5cm]{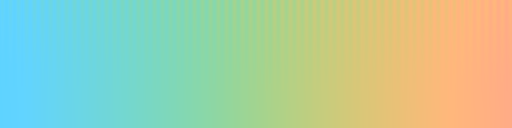}\\ 
Isoluminant colour maps are constructed from colours of equal
perceptual lightness.  These colour maps are designed for use with
relief shading.  On their own these colour maps are not very useful
because features in the data are very hard to discern.  However, when
used in conjunction with relief shading their constant lightness means
that the colour map does not induce an independent shading pattern
that will interfere with the structures induced by the relief shading.
The relief shading provides the structural information and the colours
provide the metric, data classification information.

Colour maps may have multiple attributes.  For example,
diverging-linear or diverging-isoluminant.  In addition to isoluminant
maps one can construct low lightness contrast maps for use with relief
shading.  The aim being to combine the perceptual cues that might be
obtained from, say, a linear or diverging colour map with the
perceptual cues induced by relief shading.

\subsection{Linear Colour Maps}
\label{sub:linear_maps}
\begin{figure}
\centering
\includegraphics[width=12.cm]{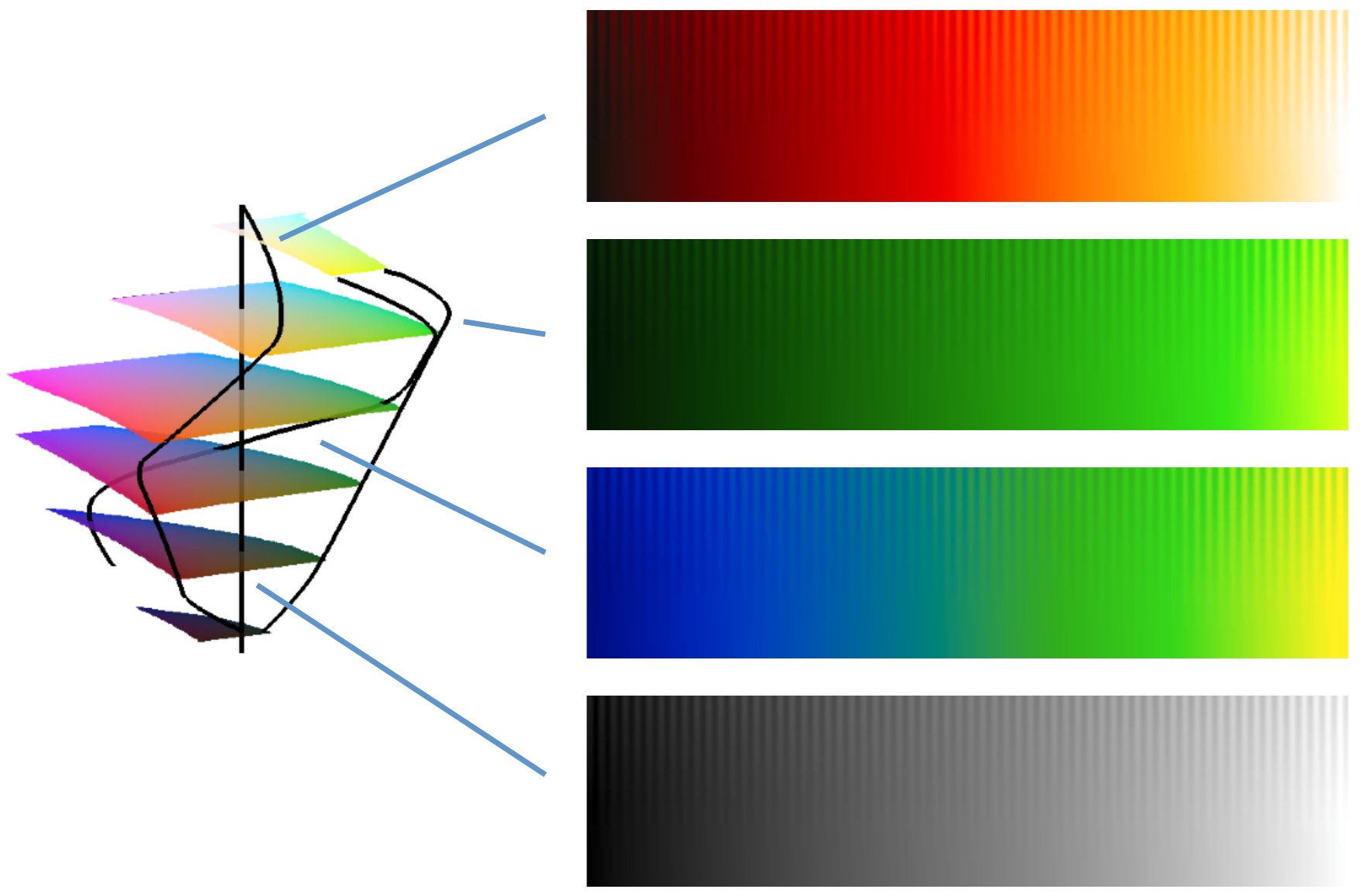}
\caption{Some linear colour maps and their paths through CIELAB space.}
  \label{fig:linear_paths}
\end{figure}

The distinguishing feature of these colour maps is that the lightness
values vary in a linear manner even though the colour map path itself
may be curved.  This linear variation of lightness, either
monotonically increasing or decreasing, induces a clear ordering of
colours making interpretation of data straightforward.  Thus, linear
colour maps are suitable for general purpose data display.  Some
examples are shown in Figure~\ref{fig:linear_paths}.

% generatepathplots.m

It can be useful to constrain the lightness values to, say, 10 to 95
rather than using the full range of 0 to 100.  It is often the case
that monitors and printers display a reduced lightness range more
reliably with features at the dark and light ends of the colour map
being less susceptible to saturation.  The overall image contrast will
be reduced slightly but the ability to identify features in the data
may be better.

\subsection{Diverging Colour Maps}
\label{sub:diverging_maps}
Diverging colour maps are intended for the display of data having a
well defined reference value where we are interested in
differentiating values that lie above, or below, this reference point.
Within the colour map the reference value is typically denoted by a
neutral colour, white, black or grey.

% generatepathplots.m
\begin{figure}
\centering
\includegraphics[width=11.5cm]{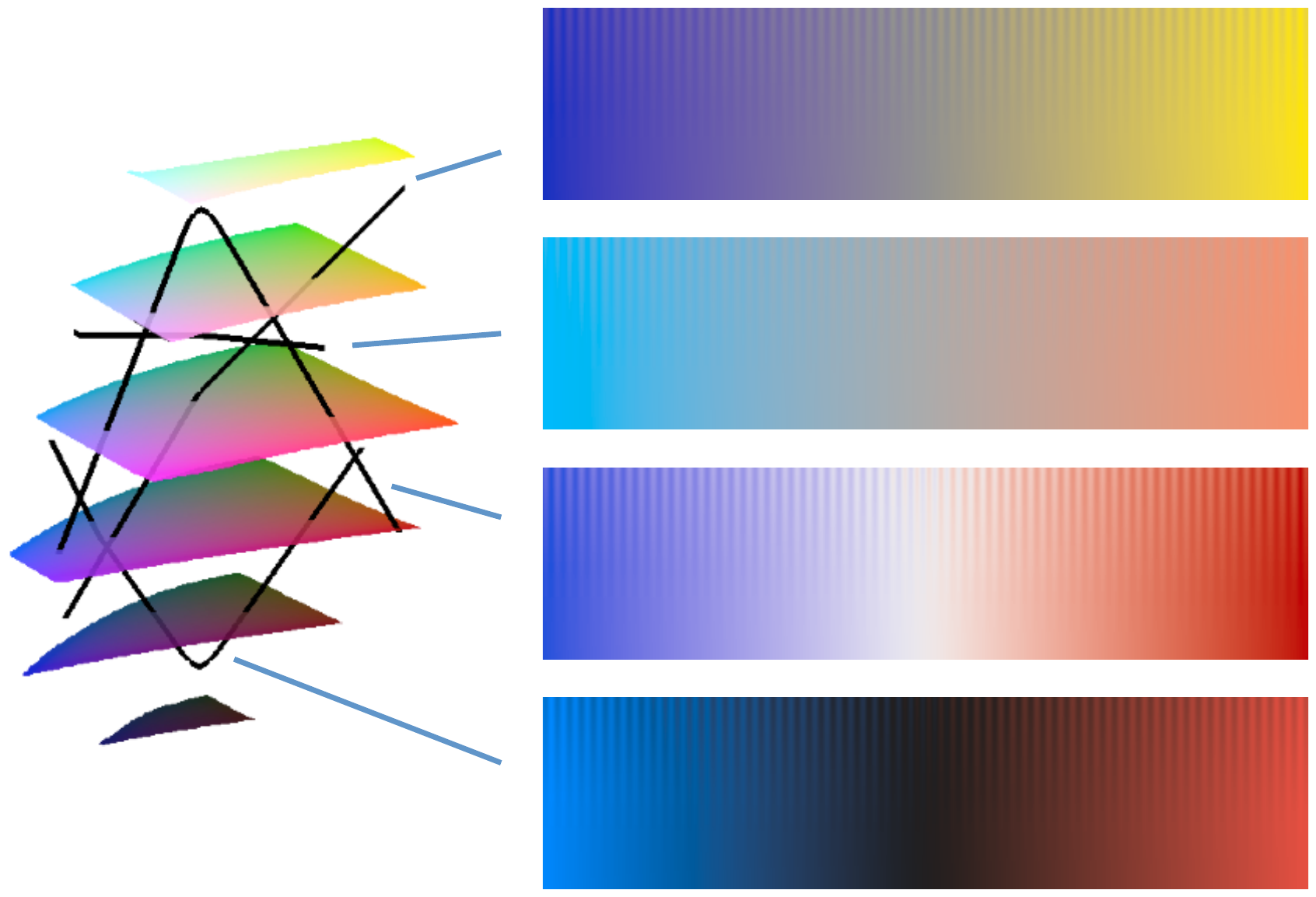}
\caption{Some diverging colour maps and their paths through CIELAB space.}
  \label{fig:diverging_paths}
\end{figure}

\begin{figure}
\centering
\includegraphics[width=8cm]{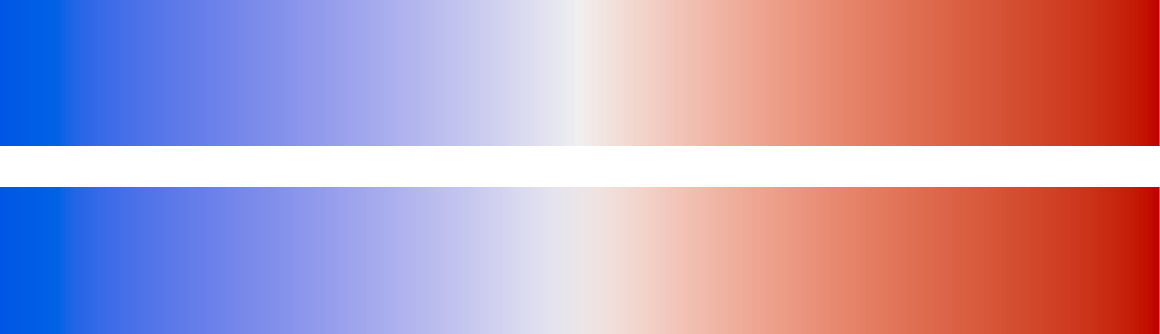}
\caption{An unsmoothed diverging colour map (top) and its smoothed version below.}
  \label{fig:diverging_smoothing}
\end{figure}

The most commonly encountered diverging colour map is a blue-white-red
map.  However, such a colour map involves a reversal of lightness
gradient at the centre.  This discontinuity in the lightness gradient
induces the perception of a false feature, see
Figure~\ref{fig:diverging_smoothing}.  To remove this the lightness
values can be smoothed, say, with a Gaussian filter.  This softens the
gradient reversal and removes the false feature.  However, a
compromise has to be accepted. The smoothing introduces a small region
in the colour map where the lightness gradient is reduced to zero.
This creates a small perceptual flat spot where structures will be
harder to see.  The degree of smoothing required is not large.  A
Gaussian filter with a standard deviation of around 5 to 7, within a
256 level colour map, is typically sufficient.  Gradient reversals in
hue or chroma prove to be relatively untroubling but smoothing of
reversals in these attributes is probably wise.

It should be noted that a lightness gradient reversal will induce a
perceptual flat spot even if no smoothing is applied. Structures in
the data with values that straddle the central reference point in the
map will be represented by colours that are effectively isoluminant,
for example, blues and reds of nearly equivalent lightness. Thus,
structures in this data range will be hard to resolve.  
Accordingly, lightness  gradient reversals in a colour map can be 
 the source of both type 1 and type 2 errors simultaneously.   Anyone 
interpreting data rendered with such a colour map should be mindful of 
this.  Thus, lightness gradient reversals in a 
colour map should be avoided where possible.

\begin{figure}
\centering
\includegraphics[width=12cm]{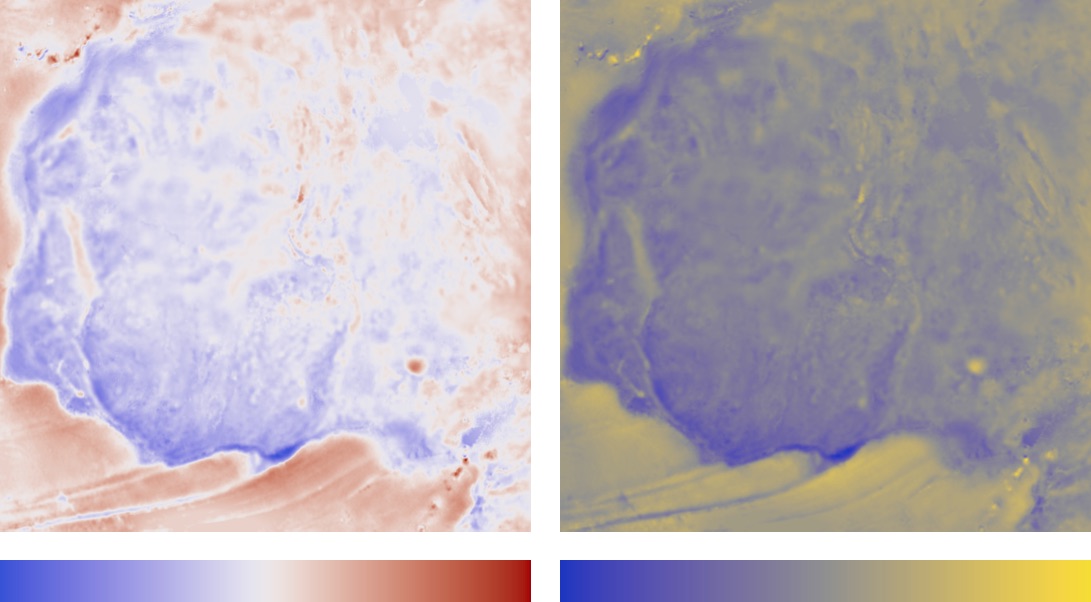}
\caption{ Residual gravity data of West Africa displayed with a
  red-white-blue colour map and a blue-grey-yellow linear-diverging
  map.  Note the easier interpretation of features near zero with the
  blue-grey-yellow map.}
  \label{fig:west_africadiverging}
\end{figure}

While a blue-white-red diverging map may be the most common other
variations are possible, as shown in Figure~\ref{fig:diverging_paths}.
For example, the reference value can be denoted by black rather than
white.  Another variation that can be very effective is a
linear-diverging map that varies from blue through grey to yellow.  By
having no lightness gradient reversal it avoids the creation of a
perceptual flat spot at the centre and provides an intuitive colour
ordering (Figure~\ref{fig:west_africadiverging}).  This kind of
diverging colour map could probably be used more widely.  One can also
conceive of an isoluminant, or low contrast, diverging colour map for
use in conjunction with relief shading.

To ensure perceptual symmetry the end points of a diverging map should
have the same chroma, and if the lightness values reverse then the end
points should also have the same level of lightness.  Without
sufficient care a diverging blue-white-red colour map constructed in
RGB space may not have this perceptual symmetry.  RGB red and blue
have lightness levels of 53 and 32 and chroma values of 105 and 134
respectively. This requirement for perceptual symmetry means that the
colour sequences that can be used for linear-diverging colour maps is
somewhat constrained by the gamut boundary.  For example, a blue-grey-yellow
sequence might not be one's first choice aesthetically (and it is against
Spence and Efendov's findings~\cite{SpenceEfendov2001}) but it allows the
maximum range of lightness values across the colour map while also
permitting end point colours with chroma that is both large and equal.

It is worth commenting here that for data to be rendered correctly
with a diverging colour map it is important that the data values are
respected so that the desired reference value within the data is
correctly associated with the central entry of the diverging colour
map.  In many visualization packages the default display methods may
not respect data values directly.  Typically, data values are
normalised by applying an offset and rescaling, before rendering with
a colour map for display.  Obviously this can lead to incorrect
display of data with a diverging colour map.

\subsection{Rainbow Colour Maps}
\label{sub:rainbow_maps}
The construction of rainbow colour maps requires a contrived path
through CIELAB space involving reversals in the lightness gradient
which can upset a viewer's perceptual ordering of the colours in the
map~\cite{RogowitzTreinish1988,Brewer1994a,Brewer1997,BorlandTaylor2007}.
Thus, rainbow colour maps are generally not recommended.  However, it
would appear unlikely that people will stop using them.  It might be
argued they have a legitimate use where the main aim is to
differentiate data values rather than communicate a data ordering.
Brewer~\cite{Brewer1997}, while cautioning against their
indiscriminate use, also accepts that rainbow maps will continue to be
used because of their attractive vibrancy.  She also makes the case
that rainbow colour maps can be used effectively as diverging maps,
using yellow to indicate the data reference point.

\begin{figure}
\centering
\includegraphics[width=12.5cm]{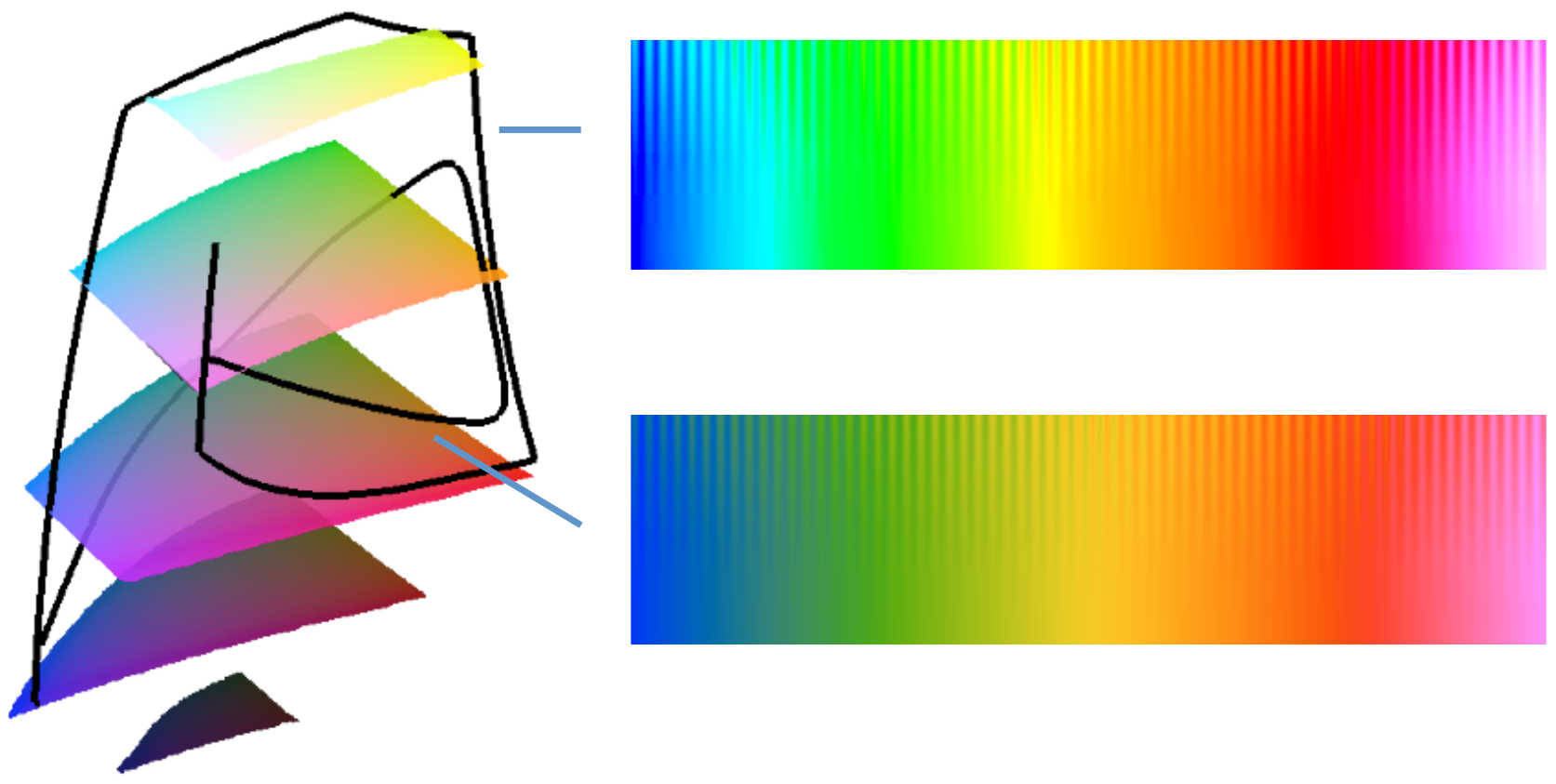}
\caption{A vendor rainbow colour map (top) and a perceptually uniform
  rainbow map (below).  The more extreme path of the vendor map
  creates false anomalies and perceptual flat spots.}
  \label{fig:rainbow_paths}
\end{figure}

With care it is possible to generate a minimally bad rainbow colour
map.  First, it is best to construct the colour map path so that in
going from blue to green it does not pass through cyan.  If cyan is
included, the subsequent colour map path from cyan through green to
yellow has very little lightness variation.  This creates an extended
region of low perceptual contrast that is not readily corrected.
False anomalies are also induced at cyan and yellow, see
Figure~\ref{fig:rainbow_paths}.  With cyan excluded, and using a less
extreme colour map path that incorporates a darker green, it is
possible to equalize the magnitude of the lightness gradient and thus
obtain uniform perceptual contrast.  If this is then followed by
smoothing of the lightness reversals at yellow and red to reduce the
perception of false anomalies at these points one can obtain a
reasonable colour map, albeit with small perceptual flat spots at
yellow and red.  See Figure~\ref{fig:rainbow_paths}.

% generatepathplots.m

\begin{figure}
\centering
\includegraphics[width=9cm]{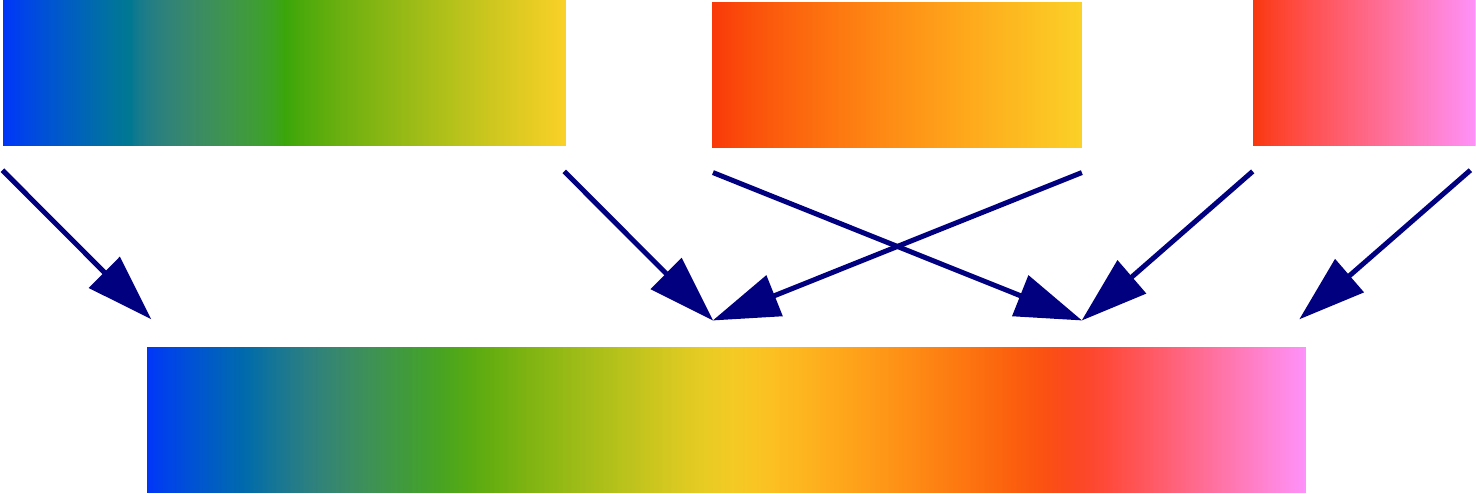}
\caption{A rainbow colour map can be thought of as the combination of
  three linear colour maps of increasing lightness.  The reversal of
  the red-yellow component creates a colour ordering inconsistency.}
  \label{fig:rainbow_reversal}
\end{figure}

The main problem with rainbow colour maps is that having yellow (and
perhaps also cyan) in the interior of the map creates a colour
ordering conflict.  The rainbow colour map presented here can be
thought of as being the combination of three linear colour maps: A
blue to yellow map; a red to yellow map; and a red to pink map.
Individually each of these colour maps provide a logical ordering of
colours with lightness values increasing from left to right.  However,
in constructing the overall colour map the red to yellow segment is
reversed when it is inserted into the composite map.  This makes the
colour ordering of the overall map inconsistent.  So while red may be
`greater than' green in terms of position in the colour map
individually the perceptual ordering of the two colours is not clear,
see Figure~\ref{fig:rainbow_reversal}.  If a rainbow colour map also
includes cyan then an additional colour ordering ambiguity is
introduced because cyan is slightly lighter than green.

\pagebreak

\subsection{Cyclic Colour Maps}
\label{sub:cyclic_maps}
\begin{figure}
\centering
\includegraphics[width=12cm]{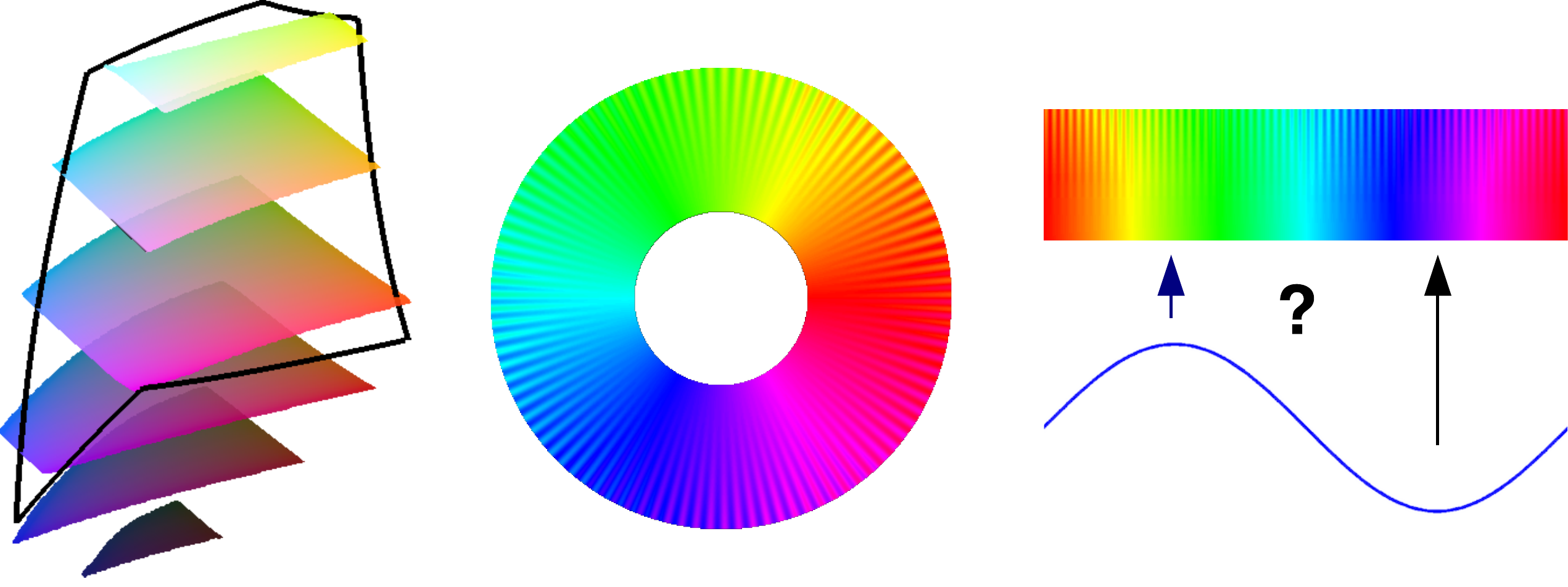}
\caption{A cyclic colour map formed from the fully saturated hue
  circle in HSV colour space and its corresponding path through CIELAB
  space. This colour map has uneven perceptual contrast and provides
  no intuitive mapping between features on a typical cyclic signal and
  the colours on the map.}
\label{fig:cyclic_hsv_problem}
\end{figure}

To present orientation values or phase data effectively in a visual
way requires the use of a cyclic colour map.  A map that is often
employed for this purpose is the hue circle from the HSV colour space.
However, this colour map has a number of problems.  The perceptual
contrast is uneven across the colour map with sections of low
lightness contrast from cyan to yellow, and from red to magenta.  The
secondary colours of cyan, magenta and yellow, being lighter, also
tend to generate false anomalies, see
Figure~\ref{fig:cyclic_hsv_problem}.  The other problem with this
colour map is that, being based on the three primary colours, it
partitions the circle into three segments; a red, a green, and a blue
segment.  These are separated by smaller sections corresponding to the
secondary colours.  This is not consistent with the way in which we
typically divide the circle.  Generally we tend to think of the four
main compass directions of north, south, east and west.  Or, if the
data is cyclic over $\pi$, we would think of the four orientation
angles of 0, 45, 90 and 135 degrees.  Alternatively we may be just
interested in a partitioning of angular phase into positive and
negative regions corresponding to the peaks and troughs of a periodic
sine wave.  Either way, the partitioning of the circle into three, or
six, segments as is done by the HSV colour map makes it a difficult
map to use when one is trying to communicate angular information in a
visual way. Ideally, in a manner similar to diverging colour maps,
where we wish to have a recognizable reference point, we would like
cyclic colour maps to have recognizable sections that can be related
to principal orientations of interest.  This, in conjunction with the
desire to have even perceptual contrast, means that designing good
cyclic colour maps is a challenging task.

% cyclicplots.m

A colour map formed from a circle at a constant level of lightness in
CIELAB space is an obvious map to construct.  However, a difficulty is
that the size of any circular path centred on the $a = 0$, $b = 0$
axis is heavily constrained by the gamut boundary.  A lightness level
of around 65 to 75 allows the largest diameter circles to be
constructed.  Even so, the maximum chroma is only about 40 and the
colours obtained are not very vivid making it hard to identify any
reference regions in the colour map.  Also, the fact that the colour
map is of constant lightness means that the resolution of fine scale
features is difficult.

To obtain a colour map with good fine scale feature resolution and
with four identifiable regions that can be associated with the main
compass directions requires a colour map path that visits four
distinct colours of high chroma that are also distributed with some
symmetry of lightness values.  One strategy to achieve this is as
follows: Two light colours of equal lightness, and two dark colours of
equal lightness are chosen.  A colour map path that alternates between
light and dark colours in a cyclic zig zag pattern is then used to
form the map.  If the perceptual contrast equalization of the colour
map only takes into account the lightness differences of the colours
then the four reference colours will end up being equally spaced in
the colour map even though the path lengths between them may be quite
different.  Finding four colours with reasonably large chroma that
form a harmonious sequence is a challenge given gamut constraints.
Another factor to consider is that an important part of having colours
that can be recognized is that they should be colours that we can
readily name~\cite{GreenArmytage2010}.  A path that has proved
successful is one that traverses blue, a darkened yellow, dark red,
pink and back to blue.  Alternatively, the darkened yellow can be
replaced with green.  This produces a map with better defined
quadrants though the colour sequence is not so harmonious. In
designing such a map one often needs to incorporate additional
intermediate control points in the path to try to equalize the width
of the four colour segments.  Finally, the four lightness gradient
reversals in the colour map need to be smoothed to avoid the creation
of false features within the map.

\begin{figure}
\centering
\includegraphics[width=9.cm]{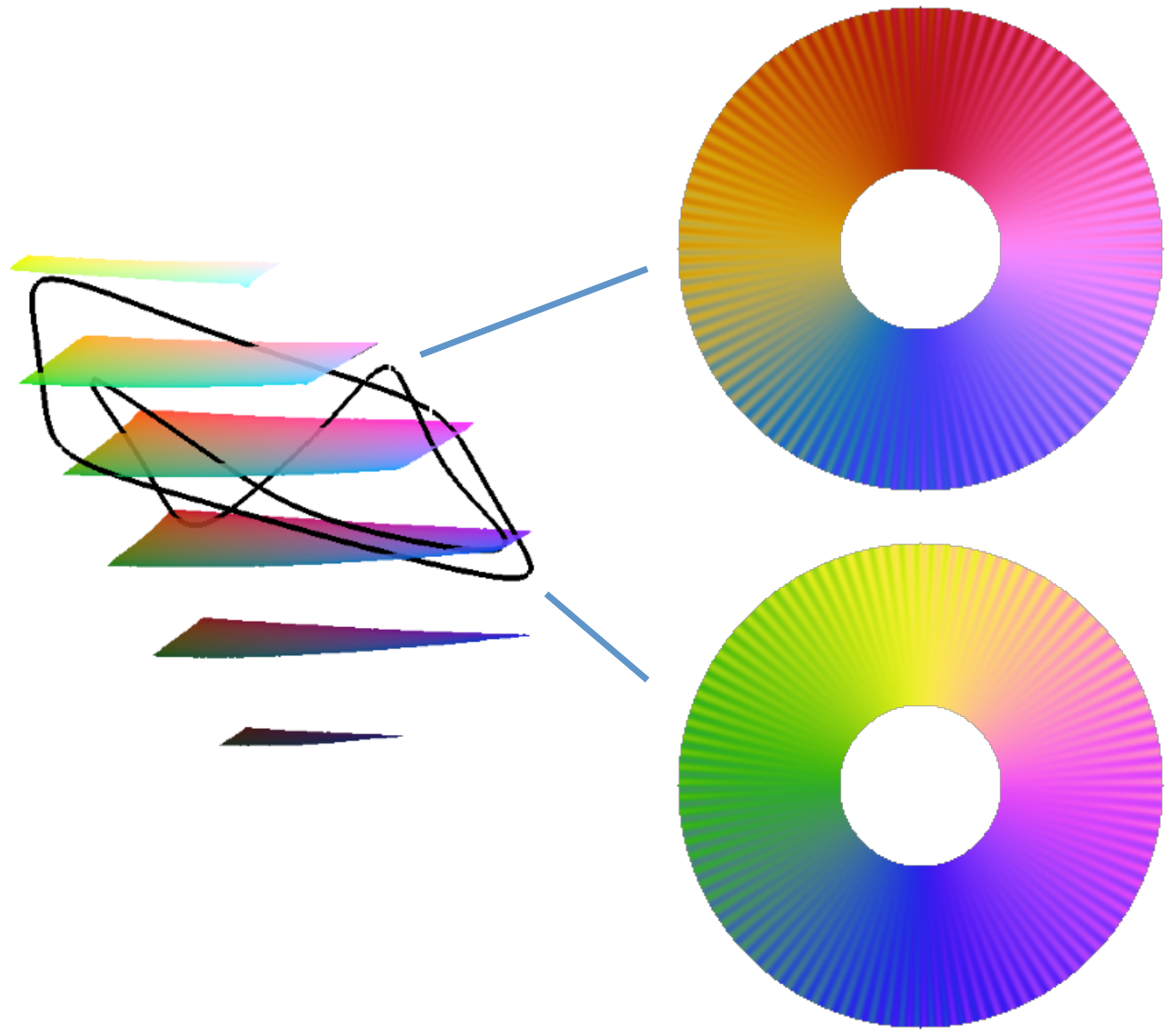}
\caption{Two cyclic colour maps designed to have four identifiable
  regions. The magenta-red-yellow-blue map is constructed from a
  cyclic zig zag path and the magenta-yellow-green-blue map is formed
  from a diamond shaped path.}
\label{fig:cyclic_fourpoints}
\end{figure}

A second strategy to achieve four identifiable regions in the map is
to form a diamond shaped path through the colour space.  A light
colour and a dark colour are chosen to form the top and bottom points
of the diamond.  Two extra colours from widely spaced locations in the
gamut at a lightness level half way between the first two colours are
used to complete the diamond shaped path.  As with the zig zag path
the fact that the sequence of colours form equal steps in lightness
difference means that the four colours will end up being equally
spaced in the colour map.  A diamond shaped path passing through
magenta, yellow green and blue is quite successful.  See
Figure~\ref{fig:cyclic_fourpoints}.
% cyclicplots.m
% cyclicplots.m
\begin{figure}
\centering
\includegraphics[width=9.cm]{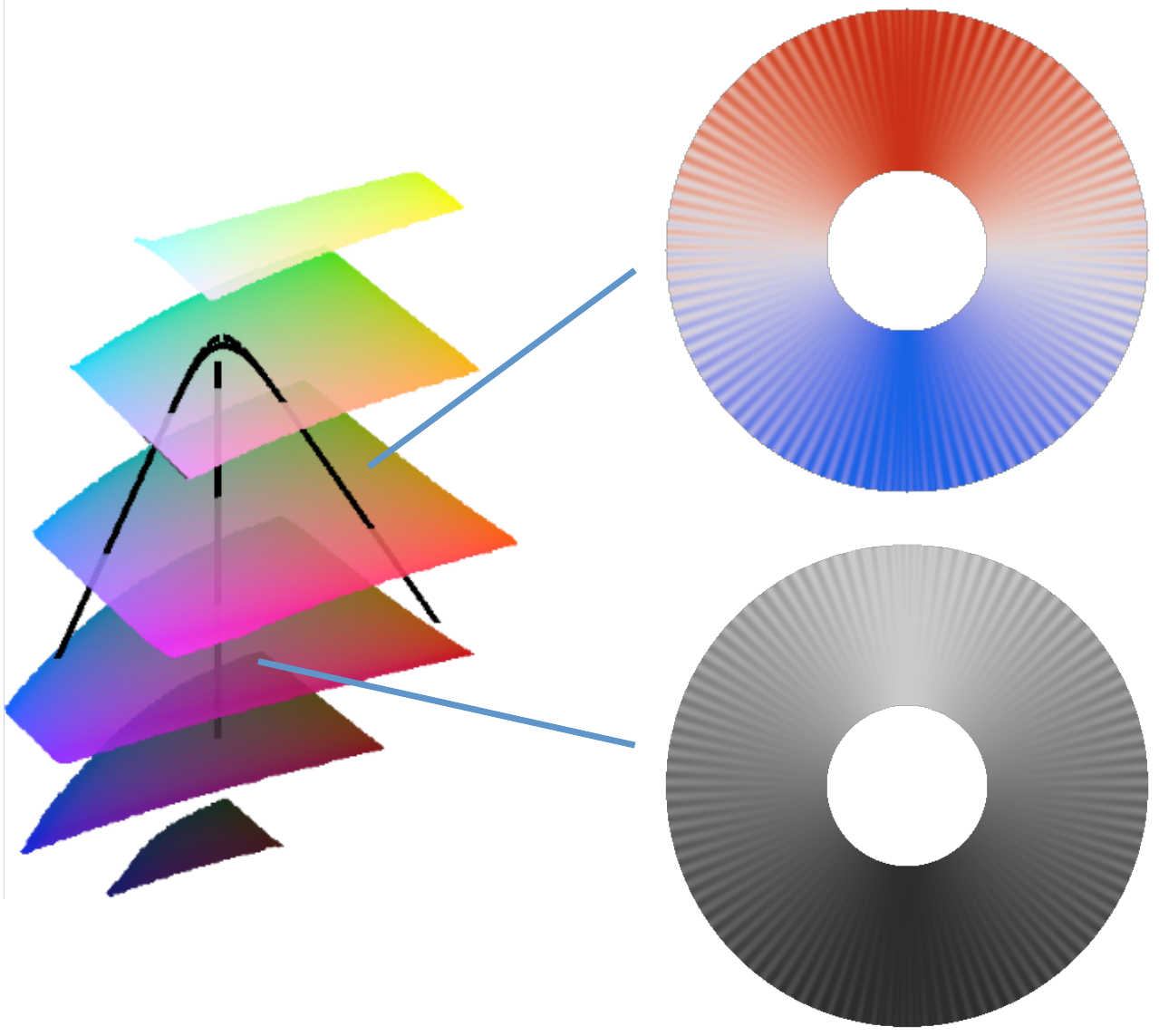}
\caption{Cyclic colour maps formed from a reversing
  white-red-white-blue-white path and from a reversing grey scale
  path.}
\label{fig:cyclic_linear}
\end{figure}

If one is prepared to accept a colour ambiguity corresponding to phase
angles of 0 and 180 degrees then the principles used for diverging
colour maps can be employed.  One can construct a cyclic map that
starts at a neutral colour, increases to a saturated colour, returns
to the neutral colour, increases to a different saturated colour
before returning to the neutral colour to complete the cycle.  A
white-red-white-blue-white colour map following this scheme is shown
in Figure~\ref{fig:cyclic_linear}.  It is also possible to form a
cyclic grey scale in a related manner.  As with diverging colour maps
smoothing of the lightness gradient reversals is required.  When
interpreting data that has been rendered with such a colour map one
has to resolve the ambiguities that occur at phase angles of 0 and 180
degrees by context.

\begin{figure}
\centering
\includegraphics[width=8.5cm]{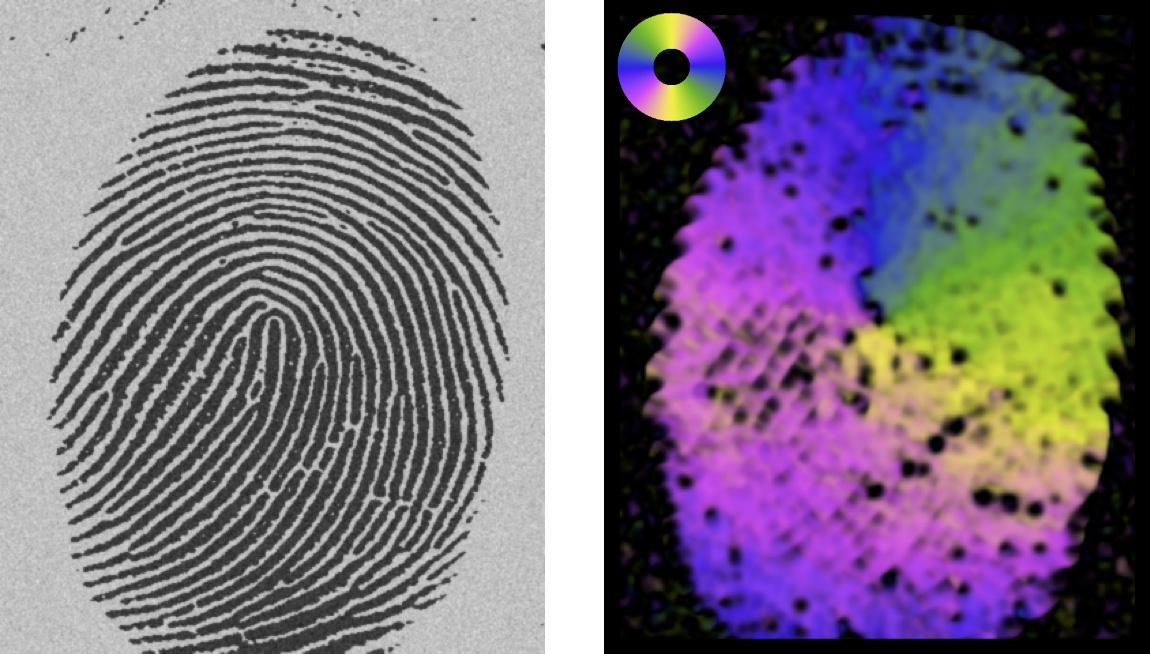}
\caption{Orientations within a fingerprint.  Note the use of the
  cyclic colour map over a cycle length of $\pi$ with the rendered
  colour values being modulated as a function of the local angular
  coherence and signal reliability. The four colour zones of the map;
  blue, magenta, yellow and green correspond to ridge orientations of
  0, 45, 90 and 135 degrees respectively.}
\label{fig:fingerorient}
\end{figure}

As with diverging colour maps, for angular data to be rendered
correctly it is important that the data values are respected so that
values are correctly associated with specific entries in a cyclic
colour map.  The assignment of colours to values also needs to respect
whether the data is cyclic over 180 degrees, as with orientation data,
or over 360 degrees as with phase data.  When rendering orientation
data it can be useful to perform a cyclic rotation of the colour map,
corresponding to 25\% of its length, so that the `low' and `high'
regions of the colour map are aligned with the horizontal and vertical
directions.  See Figure~\ref{fig:fingerorient}.

The other factor to consider when displaying angular data is that it
is also often associated with auxiliary data that might provide
information about its amplitude, reliability, or coherence.  It can be
useful to use this auxiliary data to modulate the colour map rendering
of the angular data in some manner.  The approach adopted here is to
render the angular information with a chosen colour map and then, in
RGB space, scale the colours towards black, or towards white, as a
function of the associated auxiliary data.  This
modulation/desaturation allows this auxiliary information to be
simultaneously presented and, in doing so, suppress the visualization
of angular data that is of low magnitude and/or reliability.

% scalogramplot.m
%\begin{figure}
%\centering
%\includegraphics[width=12pc]{scalogram_signal.pdf}
%\includegraphics[width=12pc]{scalogram_phase_amp.pdf}
%\caption{A signal composed of the sum of two chirps and its
%  corresponding phase-amplitude scalogram.  The phase values are
%  rendered using a white-red-white-blue cyclic colour map. The
%  amplitude information is presented simultaneously via colour
%  saturation.}
%\label{fig:scalogram}
%\end{figure}

% cyclicexamplepolts.m  (yes I cannot spell)
%\begin{figure}
%\centering
%\includegraphics[width=5cm]{ozorientfigure.pdf}
%\caption{Orientations of structures within a magnetic intensity image
%  of Western Australia.  Note the use of the cyclic colour map over a
%  cycle length of $\pi$ with the rendered colour values being
%  modulated as a function of the local local signal amplitude and the
%  local circular variance of orientations.}
%\label{fig:ozorient}
%\end{figure}

%In summary, the design of good cyclic colour maps is challenging.
%Perceptual and interpretation requirements have to be reconciled with
%colour space constraints and this cannot always be achieved to one's
%full satisfaction.  

\section{Colour Maps for Relief Shading}
\label{sec:relief_shading}
Relief shading can be an effective way of presenting data.  By
treating the data as if it is a 3D surface and generating the shading
corresponding to the surface being illuminated from some direction we
can use the eye's innate ability to interpret shading patterns to
invoke a perception of the 3D shape.  However, while interpretation of
the data's `shape' is enhanced, any sense of actual data values is
diminished because shading only depends on the surface gradient.

The use of colour in conjunction with relief shading can provide a
powerful enhancement to the perception of shape induced by the
shading. In addition, colour can also be used to convey information of
data value that is lost by relief shading.  However, if colour is
misused it is also potentially destructive to relief shading.

The main consideration when combining colour with relief shading is to
ensure that the colour map does not interfere with the perception of
features induced by the shading.  A theoretical ideal is that the
colour map be of constant lightness. The reason for this is that the
perception of features within the data is provided by the relief
shading.  If the colour map itself has a wide range of lightness
values within its colours then these may induce an independent shading
pattern that could interfere with the relief shading.  Having a colour
map of uniform lightness will ensure orthogonality between the
information induced by the colours and the information induced by the
shading.

\begin{figure}
\centering
\includegraphics[width=12cm]{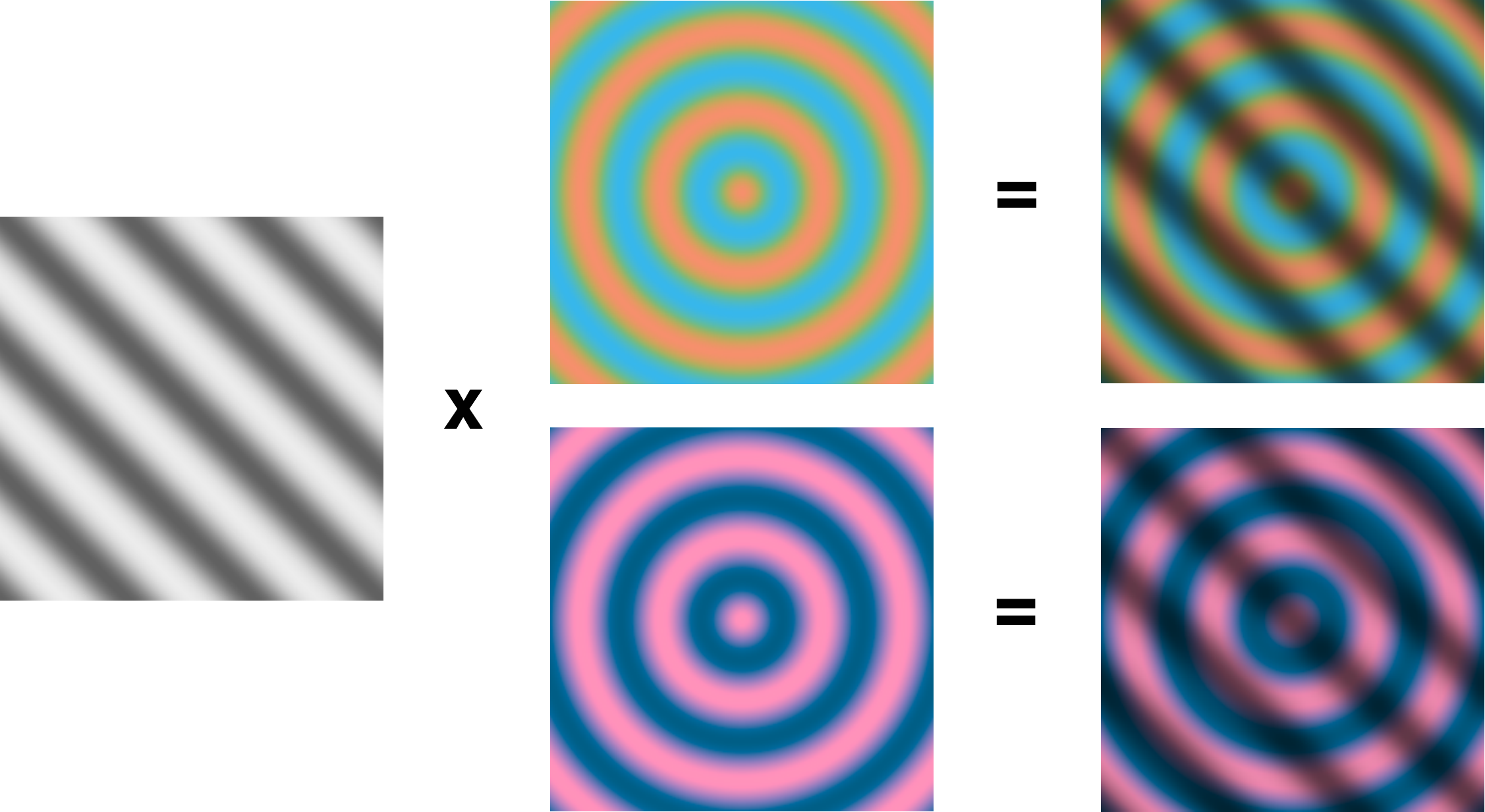}
\caption{When a shading pattern is combined with an isoluminant image
  with colour gradients that are not aligned with the shading
  gradients an amplified 3D perception of the structure is obtained
  (top right).  Combining a shading pattern with an image having
  strong lightness variations can disrupt the perception of 3D
  structure (bottom right).}
\label{fig:shading_enhancement}
\end{figure}

An interesting property of colour is that it can enhance the
perception of 3D induced by relief shading.  A constant lightness
colour map, such as the one presented in
Figure~\ref{fig:isoluminant59}, will generally produce unattractive
image renderings that are difficult to use.  However, when an
isoluminant image rendering is combined with a shading pattern such
that the colour gradients are not aligned with the shading gradients
then an amplification of the 3D shading perception can be obtained.
See Kingdom~\cite{Kingdom2003} for a description of this effect.  On
the other hand if a shading pattern is rendered with a colour map
having a significant lightness gradient the 3D structure induced by
the original shading pattern can be disrupted leading to a poor
visualization. Referring to Figure~\ref{fig:shading_enhancement} 
 notice how, in the lower-right image, that the diagonal shading
bands are no longer uniform in their darkness.  Also note that where
the darker blue regions lie alongside the shading bands the confusion
in the shading pattern is further compounded.  Compare this to the
image in the top-right of the figure and note how the shading has been
left untouched by the colouring.  Note that in this example the
shading pattern is applied to the colour image multiplicatively.  To
achieve the perception of a coloured surface being shaded the
luminance of the colours need to be modulated by the relief
shading\footnote{This is in contrast to some GIS implementations where
  it is only possible to combine a shading image with a colour image
  via a transparency blending of the two images, a weighted sum.  This
  is the wrong mechanism to use.}.

This example probably represents a worst case scenario where the
spatial frequency of the colour lightness variations is close to the
spatial frequency of the 
relief shading variations thereby maximizing the potential
interference.  While this result is interesting this synthetic example
is not typical in that the shading pattern only has a single frequency
component.  Most data sets derived from natural data have an amplitude
spectrum that decays inversely proportional to the frequency raised to
some power. That is, the amplitude spectrum is roughly proportional to
$1/f^p$ where $p$ typically ranges between 1 and
2~\cite{BakTangWiesenfeld1987,Field1987}.  If the relief shading
pattern has a more distributed frequency spectrum of this form it
appears that the colour-shading interaction effects that we see on a
simple sinusoidal shading pattern are not necessarily so strongly
reproduced.

\begin{figure}
\centering
\includegraphics[width=11cm]{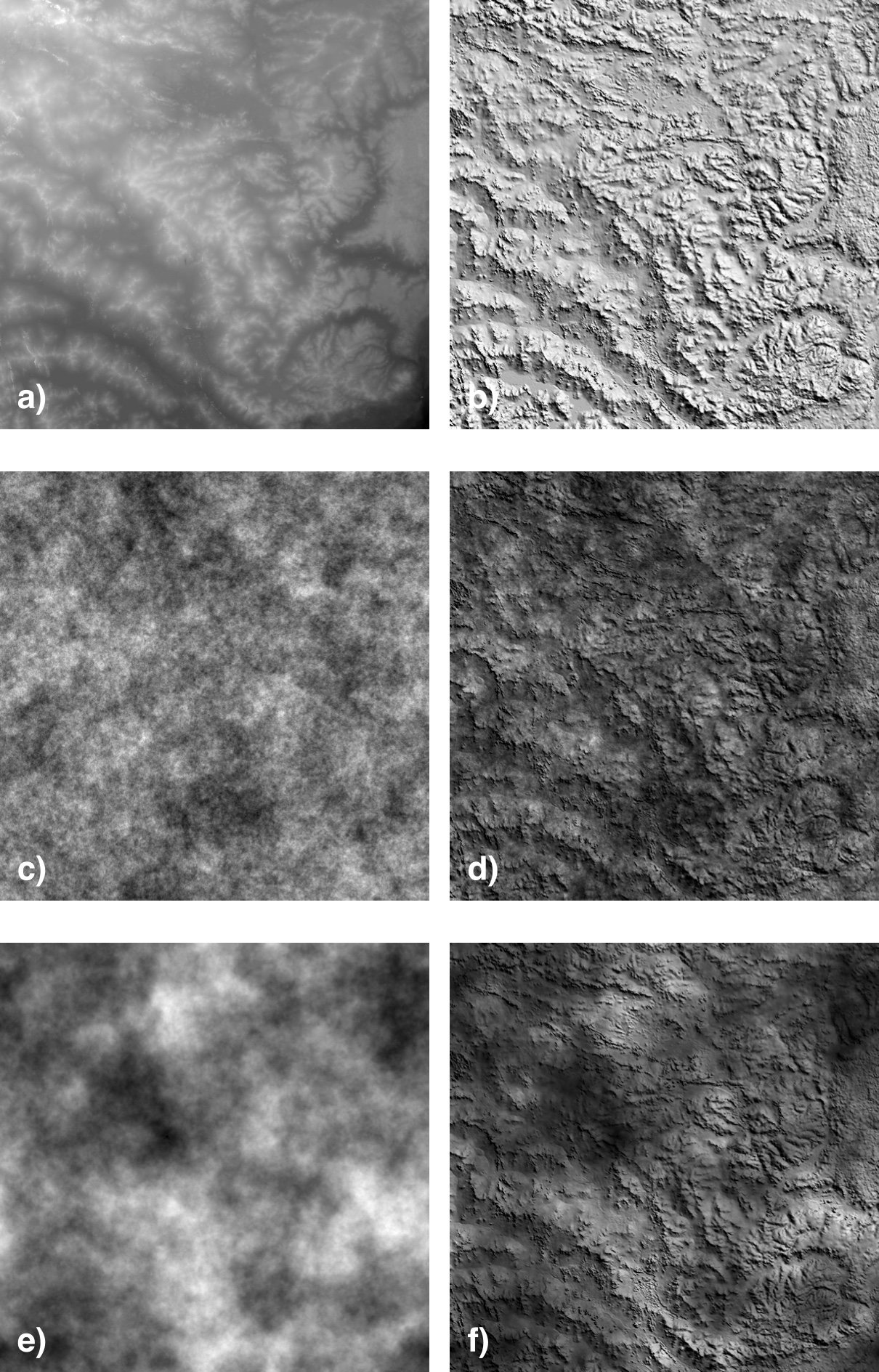}
\caption{(a) DEM image. (b) Raw relief shading of DEM.  (c)
  $1/f^{1.2}$ noise image and (d) shading rendered with the noise
  image.  Note the disruption of the shading pattern.  (e)
  $1/f^{1.8}$ noise image and (f) shading rendered with the noise
  image.}
\label{fig:one_on_f_noise}
\end{figure}

Figure~\ref{fig:one_on_f_noise} shows an example of a Digital Elevation Model (DEM) and a raw
relief shading of the data.  The amplitude spectrum of this particular
relief shading image falls off at a rate roughly proportional to
$1/f^{1.2}$.  To test how the frequency content of an overlaying image
might interfere with the 3D shape perception the shading pattern was
combined with a number of noise images having an amplitude spectrum of
the form $1/f^p$.  The noise images were grey to maximise any possible
interference with the relief shading.

As can be seen in Figure~\ref{fig:one_on_f_noise} setting $p = 1.2$,
which roughly matches the noise amplitude spectrum to the spectrum of
the relief shading, results in considerable disruption to the 3D shape
perception.  However, for $p = 1.8$, which results in a noise spectrum
more dominated by lower frequencies, there is almost no interference
with the shape perception (ignoring the regions where the level of
darkness in some of the noise patches masks the shading).  Also, not
shown for space reasons, if the shading image is combined with noise
image having a flatter spectrum using, say, $p = 0.6$ there is also
little interference with the overall 3D shape perception, though the
finer details of the shading are masked to some degree by the higher
frequency content of the noise.

%\begin{figure}  % DEM was here

Empirically it appears that as long as the image that is combined with
the relief shading is not closely matched to the frequency spectrum of
the relief shading then there is no special need to employ an
isoluminant colour map. This is especially so if the image being
combined with the shading pattern is of predominantly lower
frequencies.  However, obviously, if one uses a non-isoluminant colour
map it should not have any very dark sections that could completely
mask the relief shading altogether.  This also depends, of course, on
the scaling of gradient values used to generate the shading.

If the $1/f^{1.2}$ noise image from
Figure~\ref{fig:one_on_f_noise}~(c) is rendered with an isoluminant
colour map and combined with the relief shading then, as expected,
there is no disruption to the 3D perception.  However, any apparent
amplification of the 3D perception, as was the case with the simple
sine wave shading example, appears to be very limited if indeed there
is any at all.  It would appear that once the relief shading pattern
is, in some sense, rich enough the addition of colour makes little
difference. %, see Figure~\ref{fig:isoluminant_dem}.

%\begin{figure}
%\centering
%\includegraphics[width=9.5pc]{dem_oneonfnoise_59.png}
%\caption{DEM shading pattern from Figure~\ref{fig:one_on_f_noise}~(b)
%  combined with the noise image from
%  Figure~\ref{fig:one_on_f_noise}~(c) rendered with a constant
%  lightness colour map.}
%\label{fig:isoluminant_dem}
%\end{figure}

\begin{figure}
\centering
\includegraphics[width=7cm]{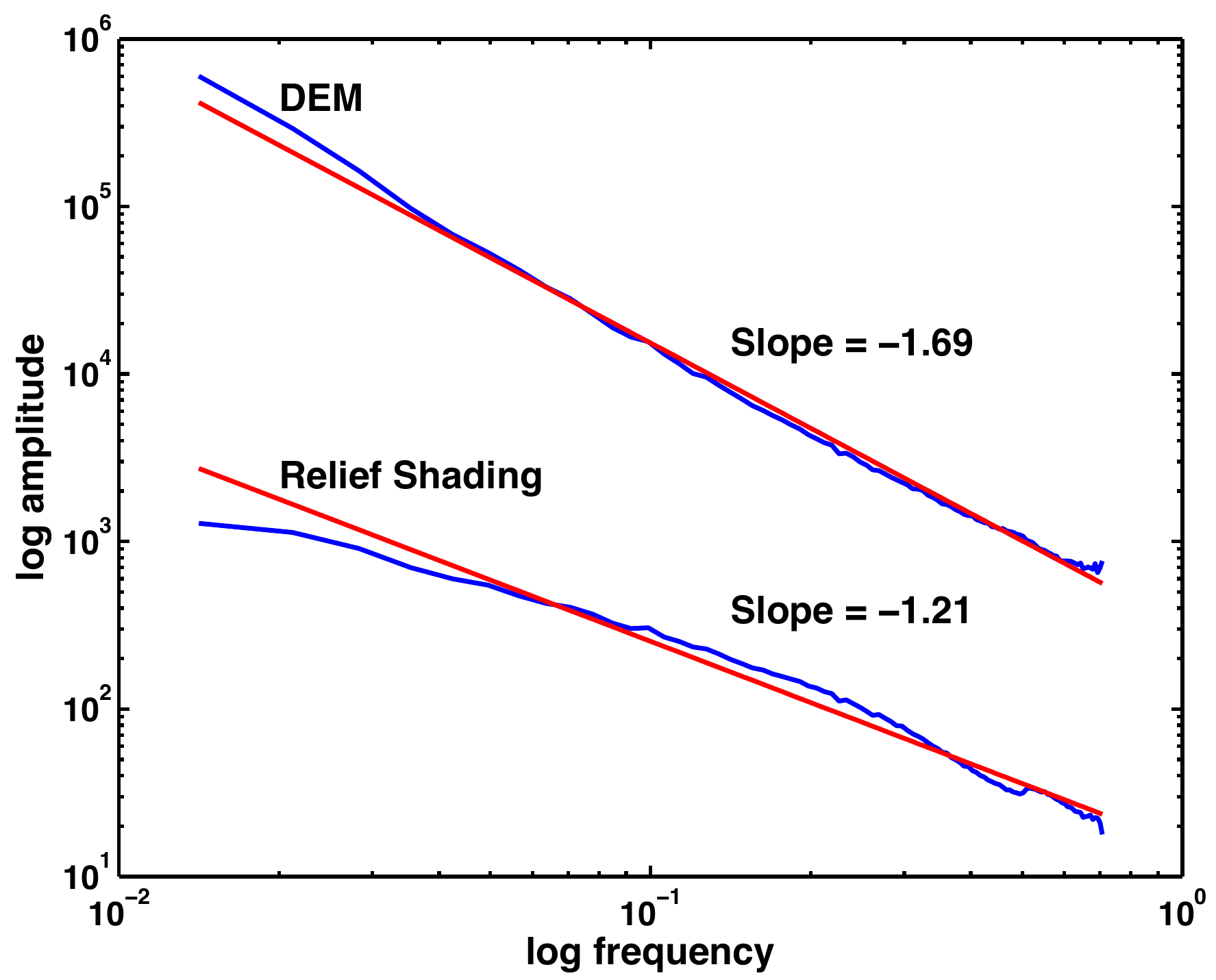}
\caption{Log-log plot of the amplitude spectrum of the DEM and its
  relief shading image.  The lines of best fit, from which the slopes
  are derived, are plotted in red.}
\label{fig:dem_relief_spectra}
\end{figure}

It is common to drape a colour image that has been derived from the
data itself over the relief shading.  The results presented here would
indicate that this practice is valid and unlikely to create any
perceptual problems.  Forming a relief image from a data set is
somewhat akin to taking a derivative of the surface.  This has the
effect of amplifying the spectral content of the image as a function
of frequency.  Thus the original data will have a stronger low
frequency content than the relief shading image.  In the example shown
in Figure~\ref{fig:one_on_f_noise} the DEM amplitude spectrum falls
away at approximately $1/f^{1.7}$ whereas the amplitude spectrum of
the raw relief shading image falls away at approximately $1/f^{1.2}$,
see Figure~\ref{fig:dem_relief_spectra}.  This difference appears to
be more than sufficient to avoid any adverse interaction between the
two.  Another reason for expecting little interaction is that the
image gradient values (and hence shading values) will, in general, be
independent of the image data values themselves.

\begin{figure}
\centering
\includegraphics[width=12cm]{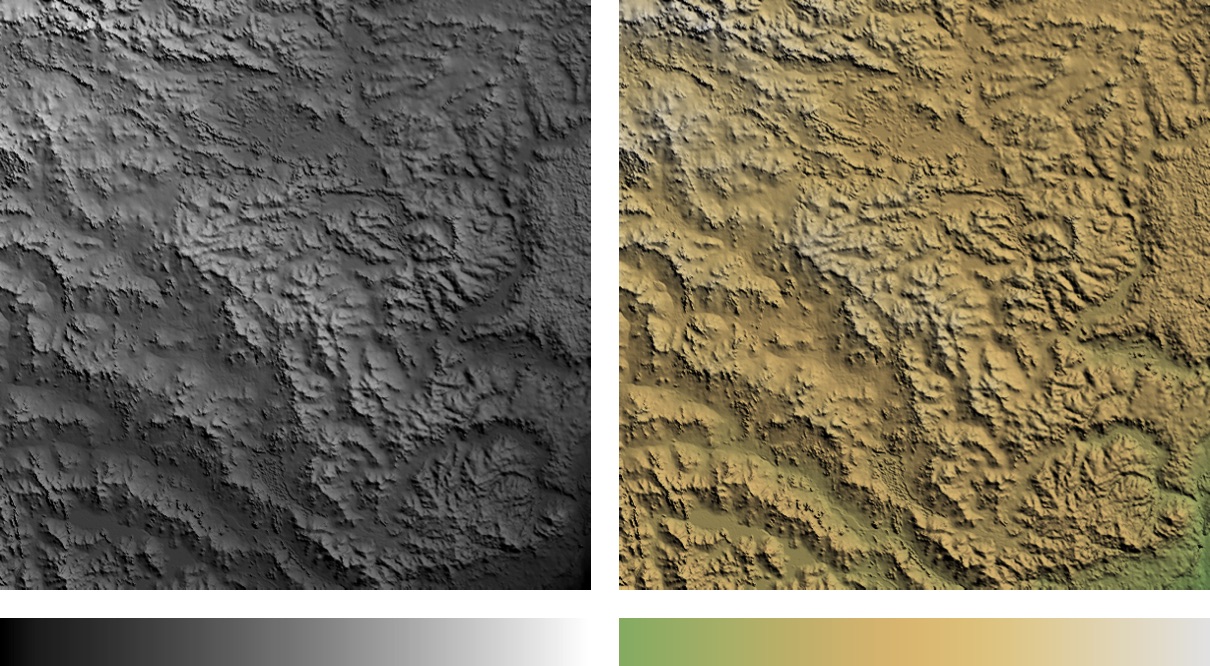}
\caption{Relief shading combined with images of the original DEM data
  rendered with a grey colour map and with a low contrast
  green-brown-white colour map.  Note that even with a grey colour map
  there is no disruption of the relief shading pattern and, in
  addition, the highest and lowest regions in the surface can still be
  identified.}
\label{fig:dem_shading}
\end{figure}

An important advantage of combining relief shading with an image that
has been derived from the data is that it allows the communication of
overall data properties, metric information, in addition to the form
and structure that is provided by the shading.  With just a raw grey
scale relief image you only get a sense of the local surface normal
information, you have no sense of the absolute data values.  If the
data range is very large this can be useful as the relief shading acts
a form of dynamic range reduction allowing small scale features to
still be seen within an arbitrarily large range of data values
(assuming that shadows are not rendered as part of the relief
shading). However, in other cases the loss of any sense of absolute
data value can be a disadvantage.  Overlaying an image derived from
the data values overcomes this problem and allows the best of both
worlds.  In the renderings of the DEM data shown in
Figure~\ref{fig:dem_shading} the highest and lowest regions in the
data can now be identified, even with the grey colour map.

\begin{figure}
\centering
\includegraphics[width=12cm]{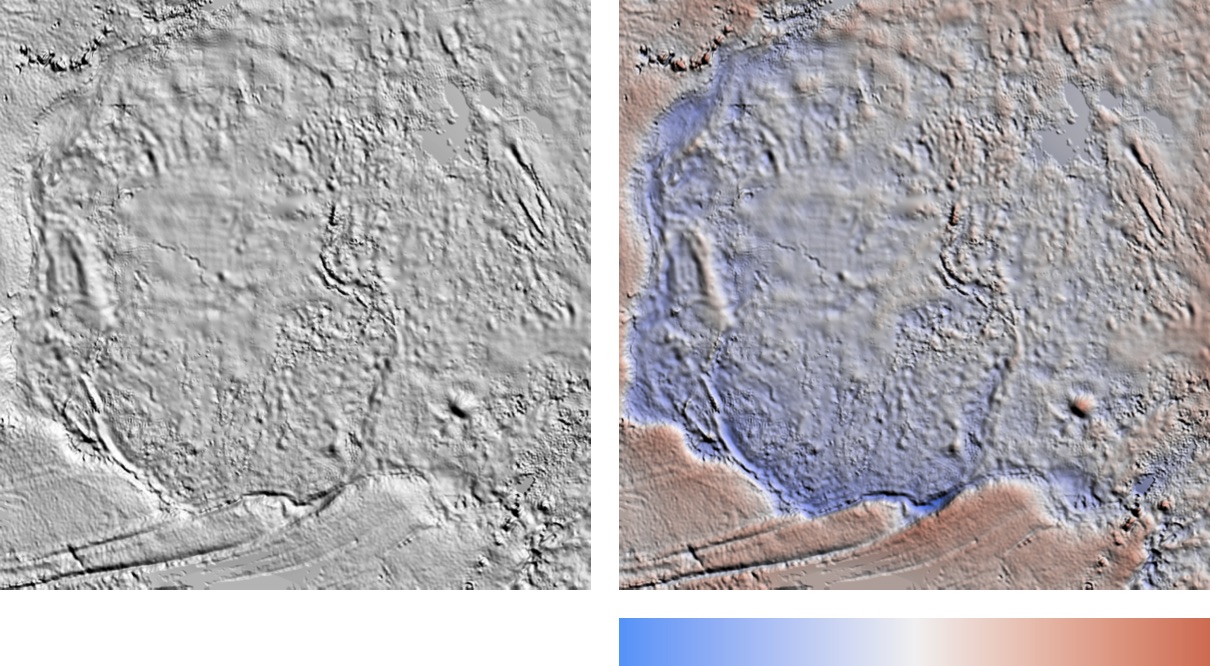}
\caption{Relief shading of residual gravity data of West
Africa and the relief shading combined with a diverging colour map
image. Note the use of a light colour map to compensate for the
darkening induced by the shading.}
\label{fig:west_africa_shading}
\end{figure}

Another example of the value of relief shading combined with a
coloured image derived from the data is shown in
Figure~\ref{fig:west_africa_shading}.  Using the residual gravity
image of West Africa shown earlier we can see that a relief shaded
image allows small scale structures to be identified readily.
However, it is hard to get a sense of the magnitude of the deviation
of features above and below zero.  Combining the relief shading with
an image of the data rendered with a diverging colour map allows the
fine structures to be seen in conjunction with the polarity of the
data.  Compare this result with the diverging colour map renderings of
this data that were presented in
Figure~\ref{fig:west_africadiverging}.

\subsection{Summary}

Relief shading combined with a coloured image, even a grey scale
image, can be a very useful way to present data.  If the frequency
content of the coloured image is significantly different from the
relief shading then no particular precautions are needed with the
colour map other than to ensure it does not have any significantly
dark colours that could mask the shading altogether.  However, as a
general guide it is probably wise to use a low contrast colour map
with a range of lightness values no more than, say, 50 to minimise any
potential disruption to the shading pattern.  In saying this one
should note that the scaling of the gradient values used to generate
the shading is arbitrary.  If the scaling of gradients is small then
sensitivity to non-isoluminance of the colour map will naturally
increase.  If the image being combined with the relief shading is
derived from another source then it is possible that the frequency
content of the two images may interfere.  Should this be the case then
an isoluminant, or very low contrast, colour map should be used.

\section{Colours for Ternary Images}

A ternary image is a colour image formed from three bands of a
multichannel image.  A basis colour is applied to each band and the
images are summed to produce the final result.  Almost invariably the
RGB primaries of red, green and blue are used as the basis colours as
this is the simplest and most obvious implementation.  Multichannel
Landsat imagery is commonly presented in this way.  Ternary images are
also used for rendering geophysical radiometric data whereby the
potassium, thorium and uranium values are used to specify the red,
green and blue components of an image respectively.

A difficulty with using the RGB primaries as basis colours is that the
perceptual sensitivity of the eye to these colours is not equal.  In
particular the eye is quite insensitive to blue, having much fewer
blue cones than green and red ones~\cite{RoordaWilliams1999}.  This is
reflected in red, green and blue having very differing CIELAB
lightness values of approximately 53, 88 and 32 respectively.
Moreover, of equal importance is the relative lightness of the
secondary colours that are obtained when the basis colours are mixed.
Cyan, magenta and yellow have lightness values of 91, 60 and 97
respectively.  If a data channel is assigned to green, or a
combination of channels are directed to the yellow or cyan secondary
colours, then these channels will be given an inappropriately large
perceptual prominence over the others.  Thus, the RGB primaries are
not the ideal basis colours for forming ternary images.   

Indeed, the early work of Tajima~\cite{Tajima1983} recognized that
using RGB to represent 3 channels of Landsat data was not ideal given
the perceptual non-uniformity of the colour space.  He proposed
mapping the Euclidean space representing the 3 channels of data into a
portion of CIELUV space.  However this approach does not acknowledge
the fact that CIELUV is only intended to be perceptually uniform at
very low spatial frequencies.

The RGB primaries were designed to allow natural colours to be
reproduced on display monitors.  However our aims are different.  We
are not wanting to reproduce colours, instead we are seeking to {\em
  create} colours that convey information.  Accordingly we want three
basis colours that can be assigned to each channel of data in a way
that allows all the information to be seen with equal perceptual
prominence.  We do not want any channel, or combination of channels,
to be treated preferentially.  To achieve this our ideal would be to
have three basis colours that are nominally red, green and blue, that
sum to white in RGB space, that are matched in lightness, and have the
same chroma.  The secondary colours resulting from these basis colours
should also be matched in lightness and chroma.  In practice this
ideal cannot be attained but, with some effort, a workable compromise
is achievable.

Designing an objective function to find a set of basis colours that
maximally satisfy these requirements via an optimization search is
complicated by the fact that we also want the gamut of colours that
can be created by the basis colours to be as large as possible.
Accordingly, a manually constrained optimization strategy was adopted.
After some experimentation the following basis colours were
constructed

{\small
\begin{tabular}{l l l}
\\
`Red'   &  RGB: [0.90  0.17  0.00]  &  CIELAB lightness 50, chroma 92 \\
`Green' &  RGB: [0.00  0.50  0.00]  &  CIELAB lightness 46, chroma 71 \\
`Blue'  &  RGB: [0.10  0.33  1.00]  &  CIELAB lightness 44, chroma 100.\\
 &
\end{tabular}
}
\\The corresponding secondary colours have coordinates\\
{\small
\begin{tabular}{l l l}
\\
`Cyan'          & RGB: [0.10  0.83  1.00]  &  CIELAB lightness 79, chroma 43\\
`Magenta'       & RGB: [1.00  0.50  1.00]  &  CIELAB lightness 72, chroma 78\\
`Yellow/Orange' & RGB: [0.90  0.67  0.00]  &  CIELAB lightness 73, chroma 77.\\
 &
\end{tabular}
}
\\The manually defined constraints were to set the `green' basis colour
to [0.0, 0.5, 0.0] and to fix the red component of the `red' basis
colour at 0.9.  The reasoning behind this choice for `green' is that
at a lightness level of around 45 this colour is the most heavily
constrained in terms of chroma.  Thus this colour has to be the
maximal `green' possible for a given lightness level.  Fixing the red
component of the `red' basis colour at 0.9 was in recognition of the
fact that in order to ensure a more uniform lightness of the secondary
colours the red component has to be reduced from 1.0, but at the same
time we want the gamut to be kept as large as possible. Finally, the
blue component of the `blue' was also fixed at 1.0. Given these
constraints, and the condition that the colours sum to white, the
unassigned green component of 0.5 and red component of 0.1 were then
distributed between the basis colours via an optimisation search in
order to minimise the lightness differences.  This resulted in the
colours given above.

With these choices the maximum lightness difference between any of the
basis colours, or between any of the secondary colours, is kept to
about 6.  However, the maximum difference in chroma is larger than one
would like at 35.  Little can be done about this due to gamut
limitations. Given the relative importance of lightness differences at
fine spatial scales this compromise was deemed acceptable.  By
imposing different manual constraints the maximum lightness difference
between the basis colours can be made arbitrarily small but the chroma
of the basis colours, and the gamut of colours that can be created
from them, become unacceptably small.

% Generated by primarytest.m 
\begin{figure}
\centering
\includegraphics[width=12cm]{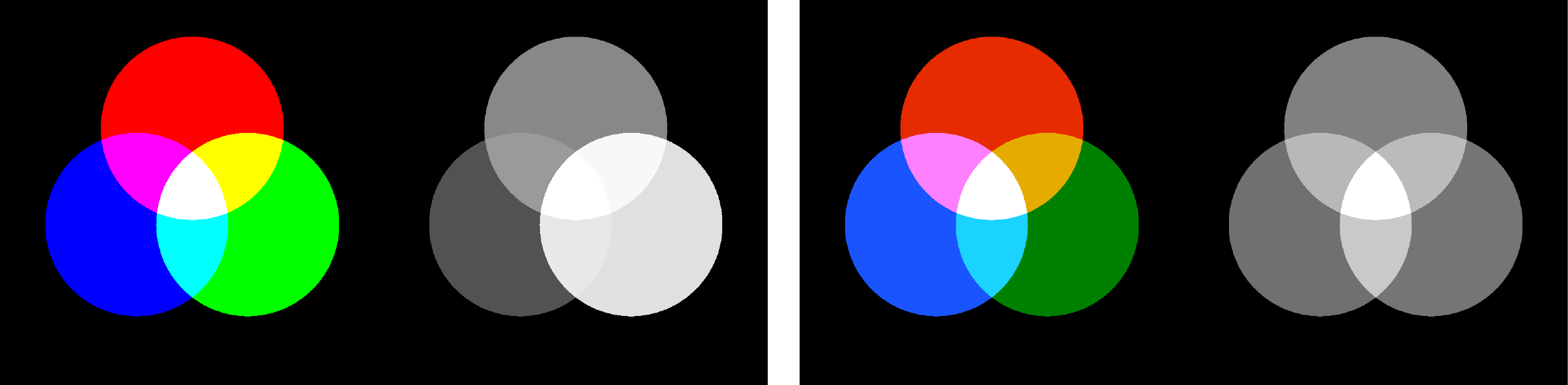}
\caption{Comparison of the RGB primaries (left) and the proposed
  ternary basis colours (right).  Also shown are the corresponding
  lightness images.  Note how the proposed basis colours are closely
  matched in lightness, as are their secondary colours.}
\label{fig:new_primaries}
\end{figure}

% Generated by isoprimaries.m 
%\begin{figure}
%\centering
%\includegraphics[width=20pc]{new_primaries_slices.pdf}
%\caption{The proposed new primary basis colours projected onto CIELAB
%  space at a lightness level of 47 (left) and the corresponding
%  secondary colours projected at a lightness level of 74 (right).  The
%  achromatic points are indicated by the red crosses.}
%\label{fig:new_primaries_slices}
%\end{figure}

\begin{figure}
\centering
\includegraphics[width=9.2cm]{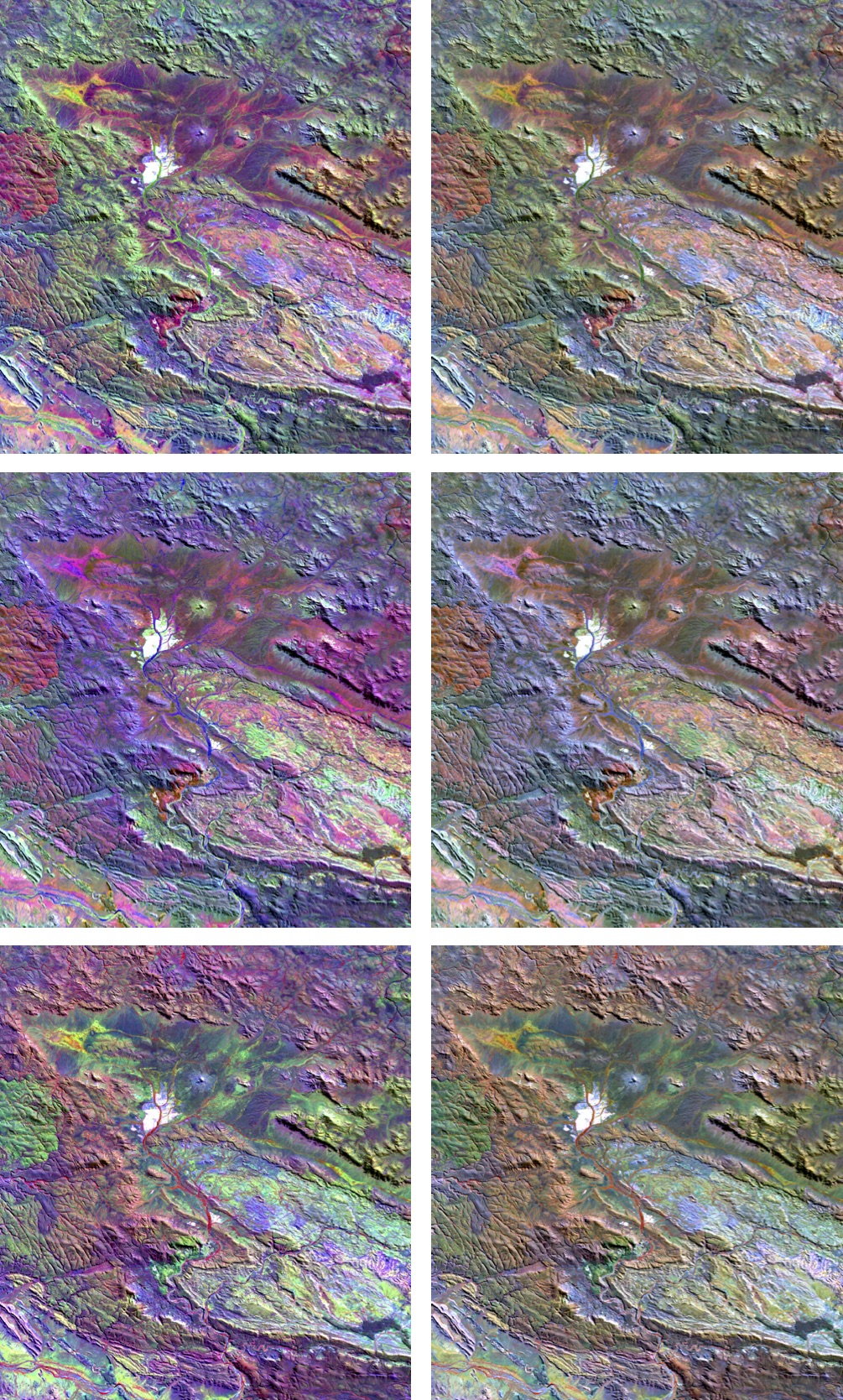}
\caption{ Left column: A series of Landsat images constructed using the
  RGB primaries as basis colours.  The images show three, of the
  possible six, permutations of colour--channel assignment.  Note how
  the structures that are most noticeable are different in each image.
  In particular, structures encoded in green are given excessive
  prominence.  Right column: An equivalent series of images constructed 
  using the proposed `red', `green' and `blue' basis colours.  The
  salience of structures is consistent across the different
  colour--channel assignments.}
\label{fig:landsat}
\end{figure}

The utility of the proposed basis colours can be evaluated by taking a
dataset with three channels and, by forming all possible permutations
of colour-channel assignment, creating six ternary images.  While
features will appear in each image with different colours we would
like the salience of the features to be consistent across the images
no matter what colour-channel permutation is used.  As shown in
Figure~\ref{fig:landsat} this is closely achieved using the proposed
basis colours.  By avoiding any bias that might arise from choosing a
particular colour-channel assignment this should allow a more
consistent and reliable interpretation of data.

An inevitable consequence of using the proposed near-isoluminant
basis colours is that the gamut of colours that can be represented is
a reduced subset of the RGB cube. This is reflected in the more
subdued colours in the ternary images shown in Figure~\ref{fig:landsat}.
However, what is gained is a more consistent representation of the
data.

\begin{figure}
\centering
\includegraphics[width=7cm]{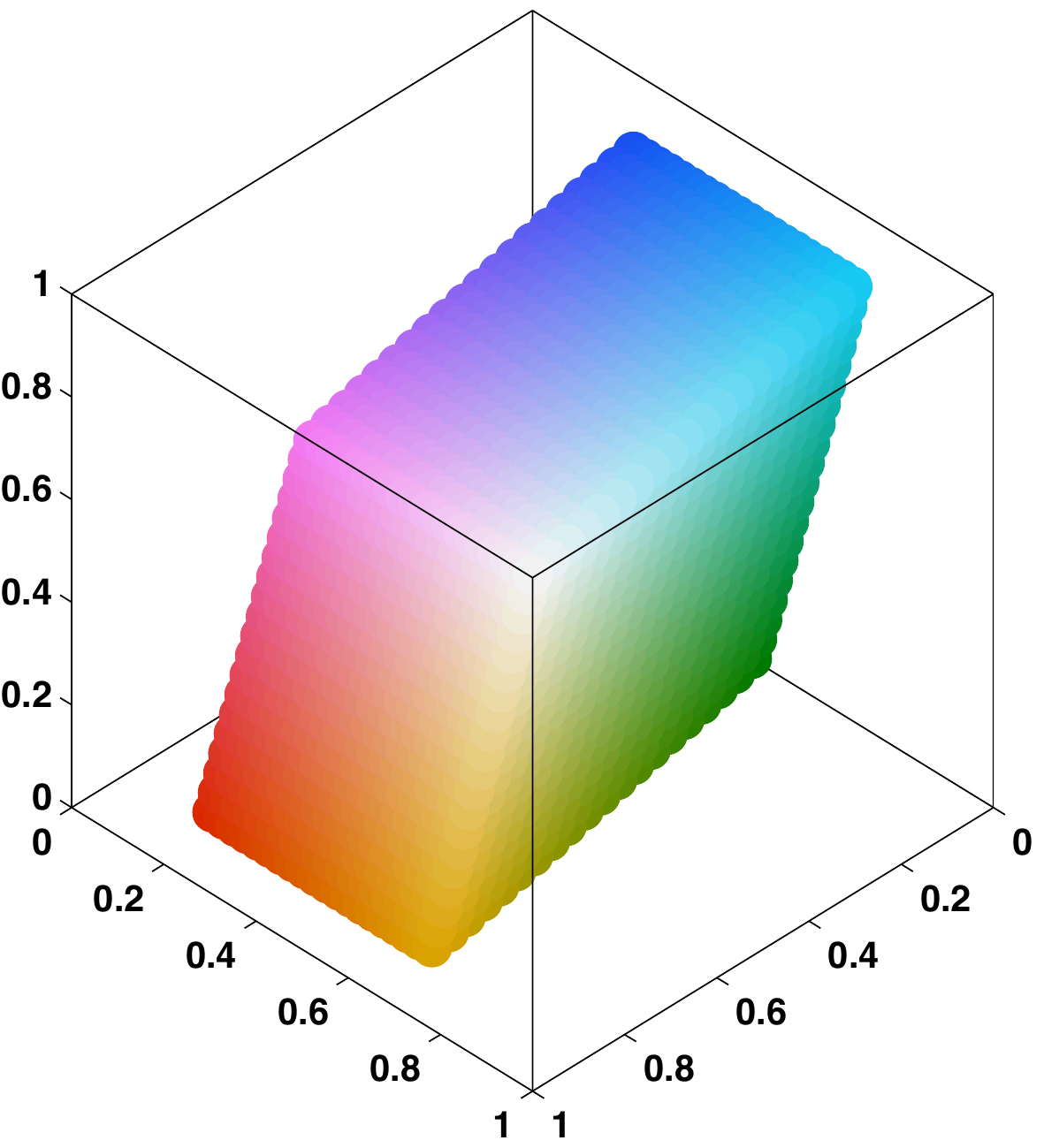}
\caption{ The gamut of colours that can be represented by the proposed
  basis colours displayed within the RGB cube.}
\label{fig:new_primaries_gamut}
\end{figure}

\pagebreak

\section{Conclusion}
\label{sec:conclusion}

This work presents a set of principled design techniques for the
construction of perceptually uniform colour maps.  Previous work in
the design of perceptually uniform colour maps has had inconsistent
success in generating good colour maps because, in many cases, there
has been a failure to recognise that CIELAB space is only designed to
be perceptually uniform at very low spatial frequencies.  The key
factor is recognising that, at the spatial frequencies that are
relevant for most image analysis tasks, it is the lightness variations
in the colour map that are most important with hue or saturation
proving to be relatively unimportant.  Once this is acknowledged then
the design process for different classes of colour maps becomes
relatively straightforward, and the compromises that have to be made
become easier to understand.

Linear colour maps are, naturally, the easiest to construct.  However
diverging, rainbow, and cyclic colour maps require lightness gradient
reversals.  These must be smoothed to avoid the creation of false
features within the colour map.  At the same time these reversals also
induce localised perceptual flat spots because the map colours in the
vicinity of these points are close to isoluminant.  This holds even if
no smoothing is performed on the lightness profile of the map.  Anyone
interpreting data with such a colour map must be mindful of this.  For
this reason linear-diverging colour maps may be better to use than the
more classical reversing ones.  Acknowledging that rainbow colour maps
will continue to be used it is shown how minimally bad rainbow maps
can be generated with reasonable perceptual uniformity even if they do
lack perceptual ordering.  Finally, the challenge of designing good
cyclic colour maps is considered and it is shown how four-colour
cyclic maps can be constructed to reflect the typical way in which we
partition the circle.

An important part of constructing good colour maps, and preventing the
use of bad ones, is the availability of simple test images that allow
colour maps to be evaluated readily.  The test images presented here
are simple to generate and have proved themselves very capable of
revealing serious deficiencies in many standard colour maps supplied by
vendors.  Hopefully the test images will also facilitate further
experimentation in colour map design strategies.

Relief shading combined with a coloured image, even a grey scale
image, can be a very useful way to present data.  This allows some
sense of metric information to be communicated in conjunction with the
form information being provided by the shading.  As long as the
frequency content of the coloured image is significantly different
from, and preferably lower than, that of the relief shading then no
particular precautions are needed with the choice of colour map.
However, low contrast maps of lighter shades will be most
effective.  This assumes though that the shading pattern has a well
distributed frequency spectrum, as is the case with a typical DEM
surface.  Where the shading pattern does not have a rich spectrum and
is primarily composed of low frequencies it may be necessary to employ an
isoluminant colour map to ensure the colours in the map do not
disrupt the perception of structure from the shading.  Under these
conditions one can expect the use of colour to amplify the perception
of structure from shading.

Finally the formation of ternary images from three basis colours is
examined in detail.  The RGB primaries are not a good choice as basis
colours and their use can result in very different perceptions of data
depending on the assignment of colours to data channels.  The
important observation is that the basis colours are required, through
a process of mixing, to create colours that communicate information in
a way that is not biased towards any channel, or combination of
channels.  Three colours that closely satisfy these needs are
presented.  They are closely matched in lightness, as are their
secondary colours, and they produce ternary images where the
perception of structures and features in the data are highly stable no
matter what colour--channel permutations are used.

MATLAB code for the design and construction of the colour maps, and
code for generating the test images is available
from:\\ \url{http://www.cet.edu.au/research-projects/colourmaps}~.\\ Also
available from this web site is a set of perceptually uniform colour
maps prepared in a variety of formats.

\section*{Acknowledgment}
This work was supported by the Centre for Exploration Targeting at
The University of Western Australia.

%\pagebreak

\appendix
\section{The Colour Map Test Image}
\label{appendix:test_image}
As shown in Figure~\ref{fig:lineartestimage} the test image consists
of a sine wave superimposed on a ramp function.  The sine wave
amplitude is set so that the range from peak to trough represents a
series of features that are 10\% of the total data range. The
amplitude of the sine wave is modulated from its full value at the top
of the image to 0 at the bottom.

%\subsection{Spatial frequency of the sine wave}
The wavelength of the sine wave is set at 8 pixels. With an image
width of 512 pixels this gives a pattern of 64 cycles.  On a computer
monitor with a nominal pixel pitch of 0.25mm this results in a
physical wavelength of 2mm.  At a monitor viewing distance of 600mm
this corresponds to 0.19 degrees of viewing angle or approximately 5.2
cycles per degree.  This falls within the range of spatial frequencies
(3--7 cycles per degree) at which most people have maximal contrast
sensitivity to a sine wave
grating~\cite{OlzakThomas1986,VanNesBouman1967}.  A wavelength of 8
pixels is also sufficient to provide a reasonable discrete
representation of a sine wave.  The aim is to present a stimulus that
is well matched to the performance of the human visual system so that
what we are primarily evaluating is the colour map's perceptual
contrast and not the visual performance of the viewer.

Inevitably, with this choice of spatial frequency, the test image is
biased towards discrimination of light-dark variations rather than
chromatic variations.  However, there is no escaping the fact that
when we look at an image we want to be able to resolve features at
this scale, and finer. 

%\subsection{Modulation of the sine wave amplitude}
The sine wave amplitude increases with the square of the distance from
the bottom of this image.  This provides a gradual rate of increase in
the amplitude with a reasonably wide, low contrast, region at the
bottom of the image allowing us to see the colour map on a ramp.  Each
row of the test image is normalised so that it spans a range of 0 to
255.  This means that the underlying ramp at the top of the image will
be reduced in slope slightly to accommodate the sine wave.

\begin{figure}
\centering
\includegraphics[width=6cm]{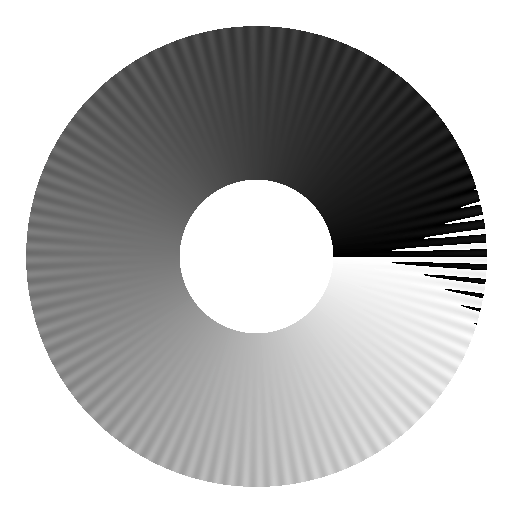}
\caption{The cyclic colour map test image rendered with a linear grey scale map.}
\label{fig:cyclic_test_image}
\end{figure}

%\subsection{Cyclic Colour Map Test Image}
An equivalent circular image for evaluating cyclic colour maps has
also been devised.  It consists of a sine wave of 100 cycles
superimposed on a spiral ramp function.  The spiral ramp starts at a
value of 0 pointing right, increasing anti-clockwise to a value of
$2\pi$ as it completes the full circle. This gives a $2\pi$
discontinuity on the right side of the image.  The amplitude of the
superimposed sine wave is modulated from its full value at the outside
of the circular pattern to 0 at the centre.  The sine wave amplitude
of $\pi/10$ means that the overall size of the sine wave from peak to
trough represents 10\% of the total spiral ramp of $2\pi$.  It is
rendered in Figure~\ref{fig:cyclic_test_image} using a linear grey
colour map to illustrate the cyclic discontinuity. 

%\pagebreak

\bibliographystyle{plain}
\bibliography{GoodColourMaps}

\end{document}